\documentstyle[aps,prl]{revtex} 
\draft 
 
\def\lsim{\mathrel{\raise.3ex\hbox{$<$\kern-.75em\lower1ex\hbox{$\sim$}}}} 
\def\gsim{\mathrel{\raise.3ex\hbox{$>$\kern-.75em\lower1ex\hbox{$\sim$}}}}

\begin{document} 
 
\twocolumn[\hsize\textwidth\columnwidth\hsize\csname
@twocolumnfalse\endcsname
 
\title {Searching for Dark Matter with Future Cosmic Positron Experiments} 
\author{Dan Hooper and Joseph Silk} 
\address{
Astrophysics Department, University of Oxford, OX1 3RH  Oxford, UK}
\date{\today} 
 
\maketitle 
 
\begin{abstract}

Dark matter particles annihilating in the Galactic halo can provide a flux of positrons potentially observable in upcoming experiments, such as PAMELA and AMS-02. We discuss the spectral features which may be associated with dark matter annihilation in the positron spectrum and assess the prospects for observing such features in future experiments. Although we focus on some specific dark matter candidates, neutralinos and Kaluza-Klein states, we carry out our study in a model independent fashion. We also revisit the positron spectrum observed by HEAT.

\end{abstract}

]

\section{Introduction}

The existence of dark matter has been confirmed by a wide array of experimental tests including observations of galactic
clusters and large scale structure \cite{structure}, supernovae
\cite{supernovae} and the Cosmic Microwave Background (CMB) anisotropies
\cite{cmb,wmap}. WMAP has measured the density of cold dark matter to be $\Omega_{\rm{CDM}} h^2 = 0.113^{+0.016}_{-0.018}$ \cite{wmap} at the 2$\sigma$ confidence level. 

For a variety of reasons, it is often thought that dark matter may be made up of weakly interacting, TeV-scale particles \cite{review}. If this is the case, there are a variety of experimental approaches which may be capable of observing them directly or indirectly. The most straight forward approach is to observe dark matter particles directly as they scatter off of a detector \cite{direct}. Such particles could also be produced and observed in colliders such as the Tevatron or the Large Hadron Collider (LHC) \cite{collider}. Alternatively, dark matter particles which annihilate in our Galaxy's halo or in gravitational wells such as the Galactic Center or the Sun may produce potentially observable fluxes of gamma-rays \cite{indirectgamma}, neutrinos \cite{indirectneutrino}, anti-protons \cite{antiprotons,ullio}, anti-deuterons \cite{ullio,Edsjo:2004pf} or positrons \cite{ullio,positrons,positrons2,positrons3,posbaltz,kkpos}.

The HEAT (High-Energy Antimatter Telescope) experiment, in three flights taking place in 1994, 1995 and 2000, observed
a flux of cosmic positrons in excess of the predicted rate, peaking around 10 GeV \cite{heat1995,heat2000,heat}. Although the source of these positrons is not known, it has been suggested in numerous publications that this signal could be the product of dark matter annihilations \cite{positrons,positrons2,positrons3,posbaltz,kkpos}.

Although HEAT provided the most accurate measurement of the cosmic positron spectrum at the time, these flights had their limitations. Due to the rapidly falling flux, HEAT was able to measure the spectrum of positrons only up to approximately 30 GeV, and with rather large error bars down to 15 GeV, or so. Fortunately, other experiments, such as PAMELA and AMS-02, are planned to improve considerably on these measurements.

PAMELA (Payload for Antimatter-Matter Exploration and Light-nuclei Astrophysics) is a satellite borne experiment designed to study the matter-antimatter asymmetry of the universe with very precise cosmic ray measurements. PAMELA's primary objectives include the measurement of the cosmic positron spectrum up to 270 GeV \cite{pamela}, well beyond the range studied by HEAT. This improvement is made possible by PAMELA's large acceptance (20.5 cm$^2$\,sr \cite{picozza}) and long exposure time (3 years). PAMELA is scheduled for launch in 2005. 

AMS-02 (Alpha Magnetic Spectrometer) is an experiment designed to be deployed on the International Space Station (ISS) for a three year mission sometime around the end of the decade, perhaps 2008. AMS-02's acceptance of about 450 cm$^2$\,sr is considerably larger than PAMELA's. This, along with superior energy resolution and electron and anti-proton rejection, makes AMS-02 the premiere experiment for measuring the cosmic positron spectrum.

The precision measurements of the cosmic positron spectrum to be provided by PAMELA and AMS-02 will be of critical importance to efforts to identify signatures of dark matter in this channel. Although the overall flux of positrons is important, without detailed spectral information it will be difficult to distinguish the products of dark matter annihilation from other possible astrophysical positron sources. Possible sources of cosmic positrons include their production via hadronic cosmic ray interactions with giant molecular clouds or electron-positron pair creation via electromagnetic interactions in nearby pulsars. Alternatively, radioactive nuclei ejected in supernova blasts could potentially produce high energy positrons \cite{heat}. 

In this article, we discuss the spectral characteristics of cosmic positrons associated with the annihilation of dark matter particles in the Galactic halo and assess the prospects for their detection by PAMELA and AMS-02. The remainder of this article is organized as follows. In section II, we discuss the spectrum of positrons produced in particle dark matter annihilations and discuss the effects of propagation on this spectrum. In section III, we show the positron spectra which result for several dark matter annihilation channels. In section IV, we discuss possible variations to these results from our choice of diffusion parameters and dark matter halo profile. In sections V and VI, we describe the positron production of two specific dark matter candidates: neutralinos in models of supersymmetry and stable Kaluza-Klein states in models of Universal Extra Dimensions (UED). In section VII, we discuss our results in the context of the measurements made by the HEAT experiment. In sections VIII and IX, we discuss the prospects for the detection of dark matter annihilation in the future experiments PAMELA and AMS-02. In section X, we compare this technique to other dark matter detection methods. Our concluding remarks are contained in section XI.


\section{Positron Production and Propagation}

Positrons can be produced in the annihilations of dark matter particles through many channels. Annihilations which yield gauge bosons, for example, yield positrons as these gauge bosons decay, peaking at an energy near $\sim m_{X}/2$. In addition to the decays $Z \rightarrow e^+ e^-$ and $W^+ \rightarrow e^+ \nu$, positrons can be produced in gauge boson decays to muons, $Z \rightarrow \mu^+ \mu^- \rightarrow e^+ \nu \bar{\nu} e^- \nu \bar{\nu}$ or $W^+ \rightarrow \mu^+ \nu \rightarrow e^+ \nu \bar{\nu} \nu$. Gauge bosons which decay to tau pairs can produce positrons directly in their decays, through their decays to muons, or hadronically through the decay of charged pions. Of course, gauge bosons may also decay to hadrons which also produce positrons via charged pions.

Dark matter annihilations to bottom quark pairs are common, especially in the case of bino-like, neutralino dark matter. These b quarks decay, producing charged leptons without the feature seen in gauge boson decays near $\sim m_{X}/2$. Well below this energy, the positron spectrum from b decays is similar to that for gauge boson decays.

\begin{figure}[thb]
\vbox{\kern2.4in\includegraphics{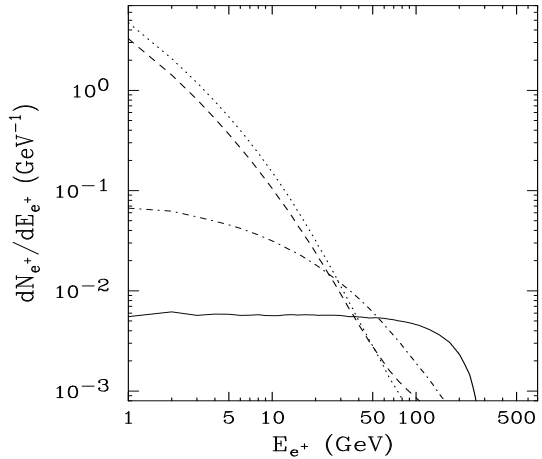}}
\caption{The positron spectrum from dark matter annihilations prior to propagation for several annihilation channels. Dotted and dashed lines represent the positron spectrum, per annihilation, for annihilations to $b \bar{b}$ and gauge boson pairs, respectively. The dot-dashed and solid lines represent annihilations to $\tau^+ \tau^-$ and $\mu^+ \mu^-$, respectively, which produce considerably harder spectra. The spectrum for annihilations to $e^+ e^-$ is not shown, but is simply a delta function at the energy equal to the WIMP mass. For all cases shown, a WIMP mass of 300 GeV was used. This figure originally appeared in Ref.~[17].}
\label{spectrum}
\end{figure}

Dark matter annihilations to top quarks always produce a pair of bottom quarks and $W^{\pm}$ bosons, which can each generate positrons as described above. As would be expected, annihilations directly to charged leptons are capable of producing a harder spectrum of positrons than the other modes one could consider. In figure~\ref{spectrum}, we compare the positron spectrum from particle annihilations via several channels. Annihilations to gauge bosons or heavy quarks each produce a much softer spectrum than the $\tau^+ \tau^-$ or $\mu^+ \mu^-$ channels. Although it is not shown, annihilations directly to $e^+ e^-$ would be represented simply by a delta function at an energy equal to the WIMP mass. For the results shown in figure~\ref{spectrum}, and throughout this paper, the spectrum from each annihilation channel, including cascading, is calculated using PYTHIA \cite{pythia}, as it is implemented in the DarkSusy package \cite{darksusy}.

The spectra shown in figure~\ref{spectrum} are not those observed at Earth, however. As positrons travel through the Galactic halo, they move under the influence of the tangled interstellar magnetic fields and lose energy via inverse Compton and synchrotron processes. We can take these processes into account by solving the diffusion-loss equation:
\begin{eqnarray}
\frac{\partial}{\partial t}\frac{dn_{e^{+}}}{dE_{e^{+}}} &=& \vec{\bigtriangledown} \cdot \bigg[K(E_{e^{+}},\vec{x})  \vec{\bigtriangledown} \frac{dn_{e^{+}}}{dE_{e^{+}}} \bigg] \nonumber\\
&+& \frac{\partial}{\partial E_{e^{+}}} \bigg[b(E_{e^{+}},\vec{x})\frac{dn_{e^{+}}}{dE_{e^{+}}}  \bigg] + Q(E_{e^{+}},\vec{x}),
\label{dif}
\end{eqnarray}
where $dn_{e^{+}}/dE_{e^{+}}$ is the number density of positrons per unit energy, $K(E_{e^{+}},\vec{x})$ is the diffusion constant, $b(E_{e^{+}},\vec{x})$ is the rate of energy loss and $Q(E_{e^{+}},\vec{x})$ is the source term.

We parameterize the diffusion constant \cite{diffusion} by
\begin{equation}
K(E_{e^{+}}) = 3.3 \times 10^{28} \bigg[3^{0.47} + E_{e^{+}}^{0.47} \bigg] \,\rm{cm}^2 \, \rm{s}^{-1},
\label{k}
\end{equation}

and the energy loss rate by
\begin{equation}
b(E_{e^{+}}) = 10^{-16} E_{e^{+}}^2 \,\, \rm{s}^{-1}.
\label{b}
\end{equation}
$b(E_{e^{+}})$ is the result of inverse Compton scattering on both starlight and the cosmic microwave background \cite{lossrate}. The diffusion parameters are constrained from analyzing stable nuclei in cosmic rays (primarily by fitting the boron to carbon ratio) \cite{L}. 

In equations~\ref{k} and~\ref{b}, we have dropped the dependence on location, treating these as constant within the diffusion zone. For the boundary conditions of the diffusion zone, we consider a slab of thickness $2L$, where $L$ is fit to observations to be 4 kpc \cite{diffusion,L}. Beyond the boundaries of our diffusion zone, we drop the positron density to zero (free escape boundary conditions).

Although the technique used here does not explicitly include the effects of re-acceleration, these effects are introduced though the energy dependence of the diffusion constant. The effect of re-acceleration is also not important at energies above a few GeV, where we are primarily interested. 

In addition to the propagation effects described above, as positrons approach the solar system, their interaction with the solar wind and magnetosphere can become important. These effects, called solar modulation, can be roughly parameterized using the technique of Ref.~\cite{solarmod1}. Alternatively, if one assumes that the effects of solar modulation are charge sign independent, their impact can be removed by considering the ratio of positrons to positrons plus electrons at a given energy rather than the positron flux alone. This quantity, called the {\it positron fraction}, is often used in lieu of the positron flux to minimize the uncertainties associated with modelling the impact of solar modulation.

\section{The Cosmic Positron Spectrum From Dark Matter Annihilations}
\label{cps}

\begin{figure}[thb]
\vbox{\kern2.4in\includegraphics{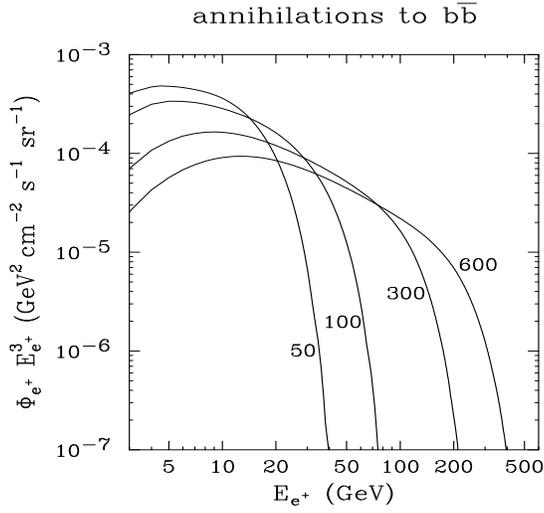}}
\caption{The spectrum of positrons, including the effects of propagation, from dark matter annihilations to b quark pairs. WIMP masses of 50, 100, 300 and 600 GeV were considered. A dark matter distribution with $BF=5$ (see section~\ref{cps}), $\rho(\rm{local})=0.43 \,\rm{GeV/cm^3}$ and an annihilation cross section of $\sigma v = 10^{-25} \, \rm{cm}^3/\rm{s}$ was used. The effects of solar modulation are not included.}
\label{fluxbb}
\end{figure}

\begin{figure}[thb]
\vbox{\kern2.4in\includegraphics{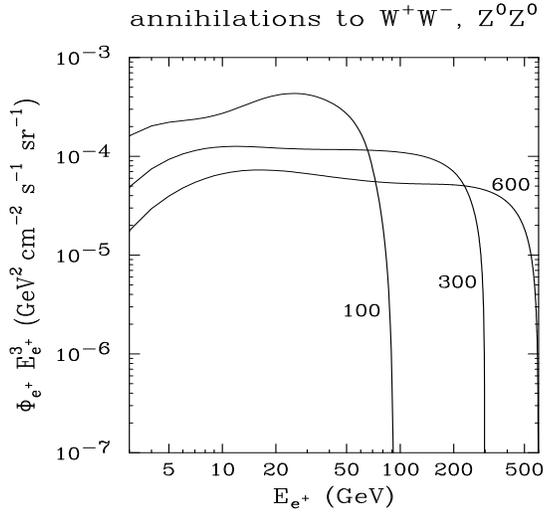}}
\caption{The spectrum of positrons, including the effects of propagation, from dark matter annihilations to $ZZ$ and $W^+ W^-$ pairs. WIMP masses of 100, 300 and 600 GeV were considered. Note the enhancement near $m_X/2$ as compared to the case of annihilations to b quarks. A dark matter distribution with $BF=5$ (see section~\ref{cps}), $\rho(\rm{local})=0.43 \,\rm{GeV/cm^3}$ and an annihilation cross section of $\sigma v = 10^{-25} \, \rm{cm}^3/\rm{s}$ was used. The effects of solar modulation are not included.}
\label{fluxgauge}
\end{figure}

\begin{figure}[thb]
\vbox{\kern2.4in\includegraphics{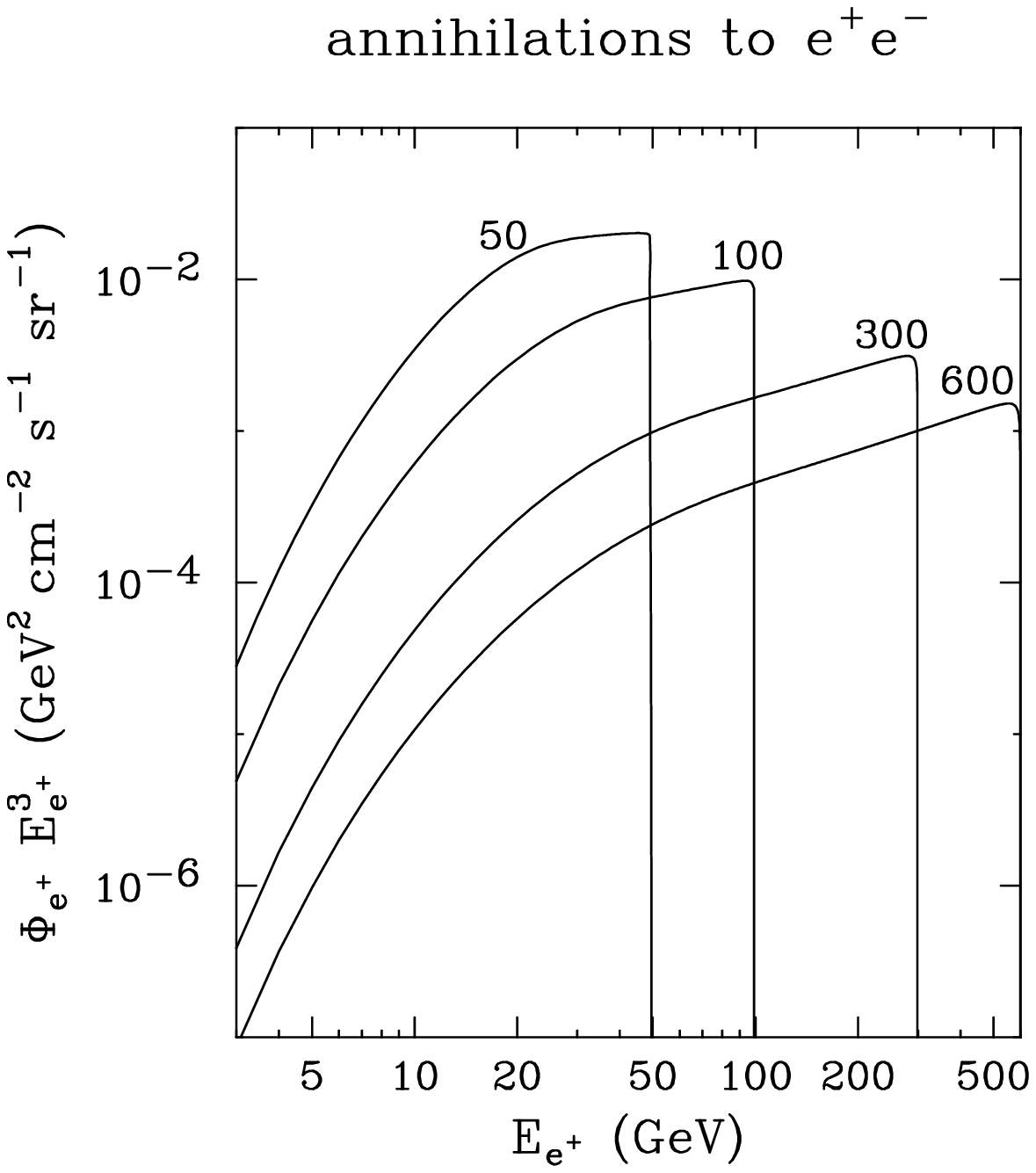}}
\caption{The spectrum of positrons, including the effects of propagation, from dark matter annihilations to $e^+ e^-$ pairs. WIMP masses of 50, 100, 300 and 600 GeV were considered. A dark matter distribution with $BF=5$ (see section~\ref{cps}), $\rho(\rm{local})=0.43 \,\rm{GeV/cm^3}$ and an annihilation cross section of $\sigma v = 10^{-26} \, \rm{cm}^3/\rm{s}$ was used. The effects of solar modulation are not included.}
\label{fluxee}
\end{figure}

\begin{figure}[thb]
\vbox{\kern2.4in\includegraphics{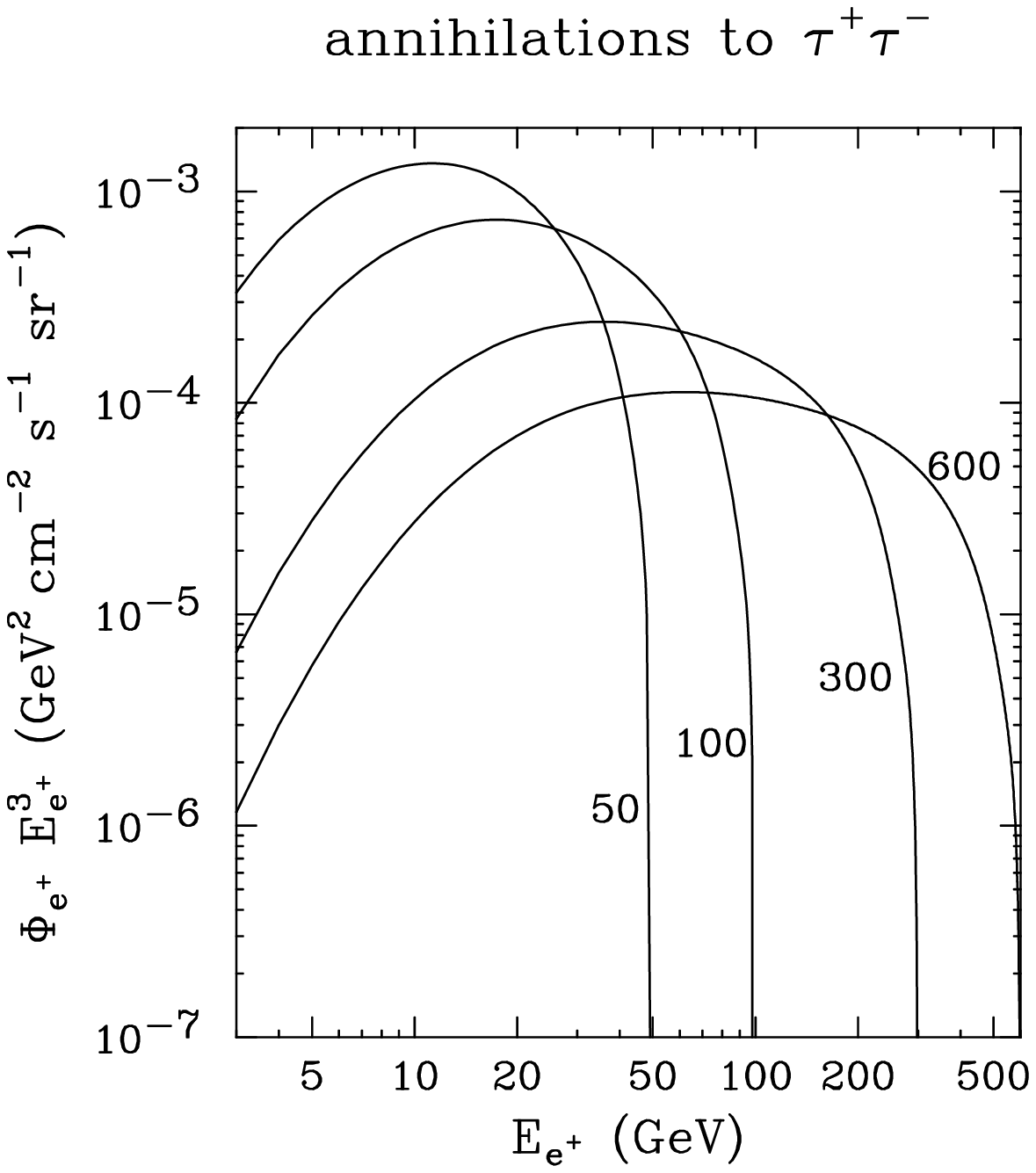}}
\caption{The spectrum of positrons, including the effects of propagation, from dark matter annihilations to $\tau^+ \tau^-$ pairs. WIMP masses of 50, 100, 300 and 600 GeV were considered. A dark matter distribution with $BF=5$ (see section~\ref{cps}), $\rho(\rm{local})=0.43 \,\rm{GeV/cm^3}$ and an annihilation cross section of $\sigma v = 10^{-25} \, \rm{cm}^3/\rm{s}$ was used. The effects of solar modulation are not included.}
\label{fluxtautau}
\end{figure}

The positron spectra produced in dark matter annihilations, after taking into account the effects of propagation, are shown for various channels in figures~\ref{fluxbb}-\ref{fluxtautau}. Many of the features described in the previous section can be identified here. In particular, the rapid decline in the spectrum at around $m_X/2$ can be seen in figure~\ref{fluxgauge}, corresponding to the energy at which direct production of positrons from gauge boson decays no longer contributes. In figure~\ref{fluxee}, an even more dramatic effect is seen at an energy equal to the WIMP mass, corresponding to the threshold for the direct production of $e^+ e^-$.

\begin{figure}[thb]
\vbox{\kern2.4in\includegraphics{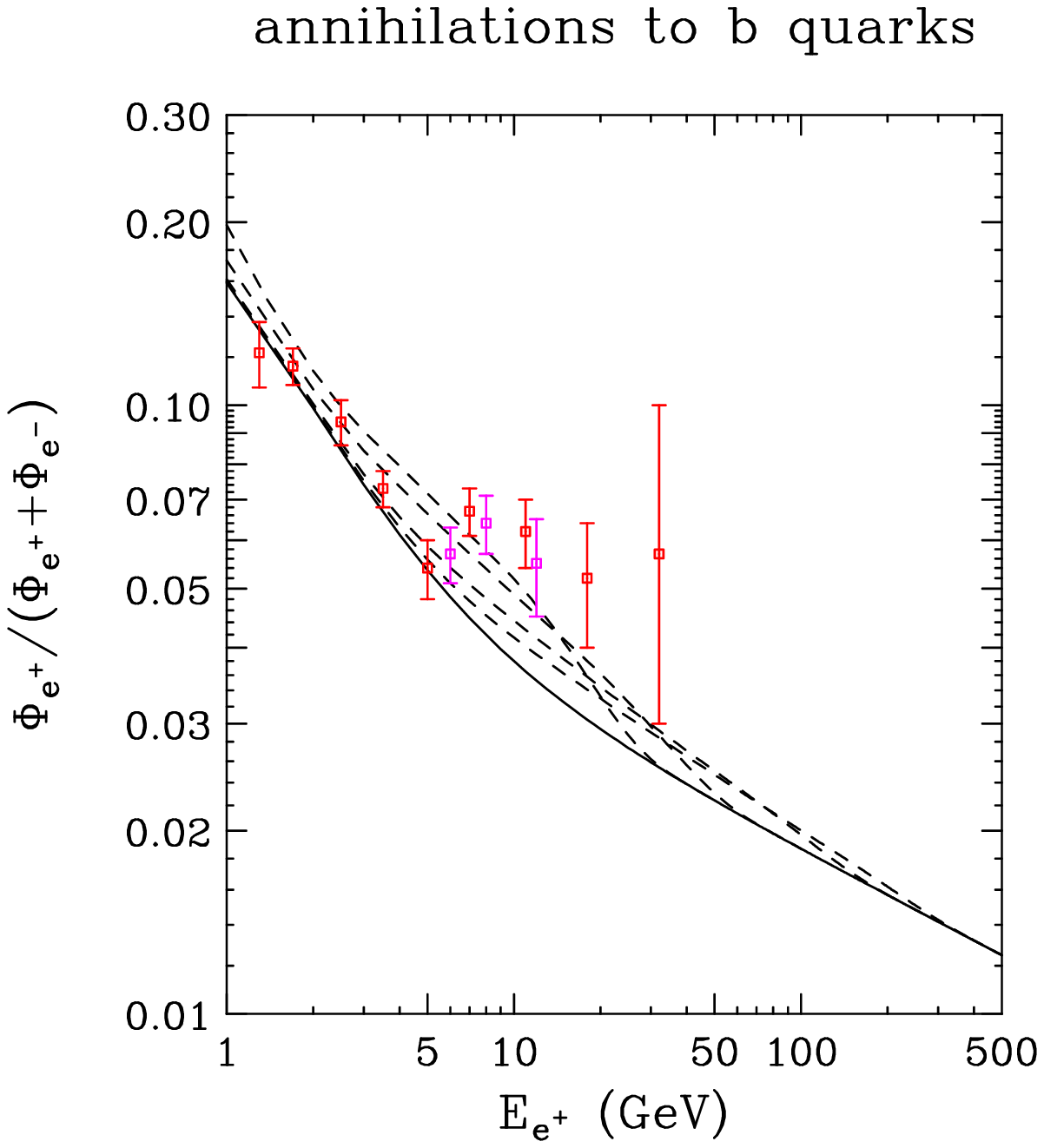}}
\caption{The positron fraction in the cosmic ray spectrum from dark matter annihilations to b quark pairs. WIMP masses of 50, 100, 300 and 600 GeV were considered. A dark matter distribution with $BF=5$ (see section~\ref{cps}), $\rho(\rm{local})=0.43 \,\rm{GeV/cm^3}$ and an annihilation cross section of $\sigma v = 10^{-25} \, \rm{cm}^3/\rm{s}$ was used.}
\label{posbb}
\end{figure}

\begin{figure}[thb]
\vbox{\kern2.4in\includegraphics{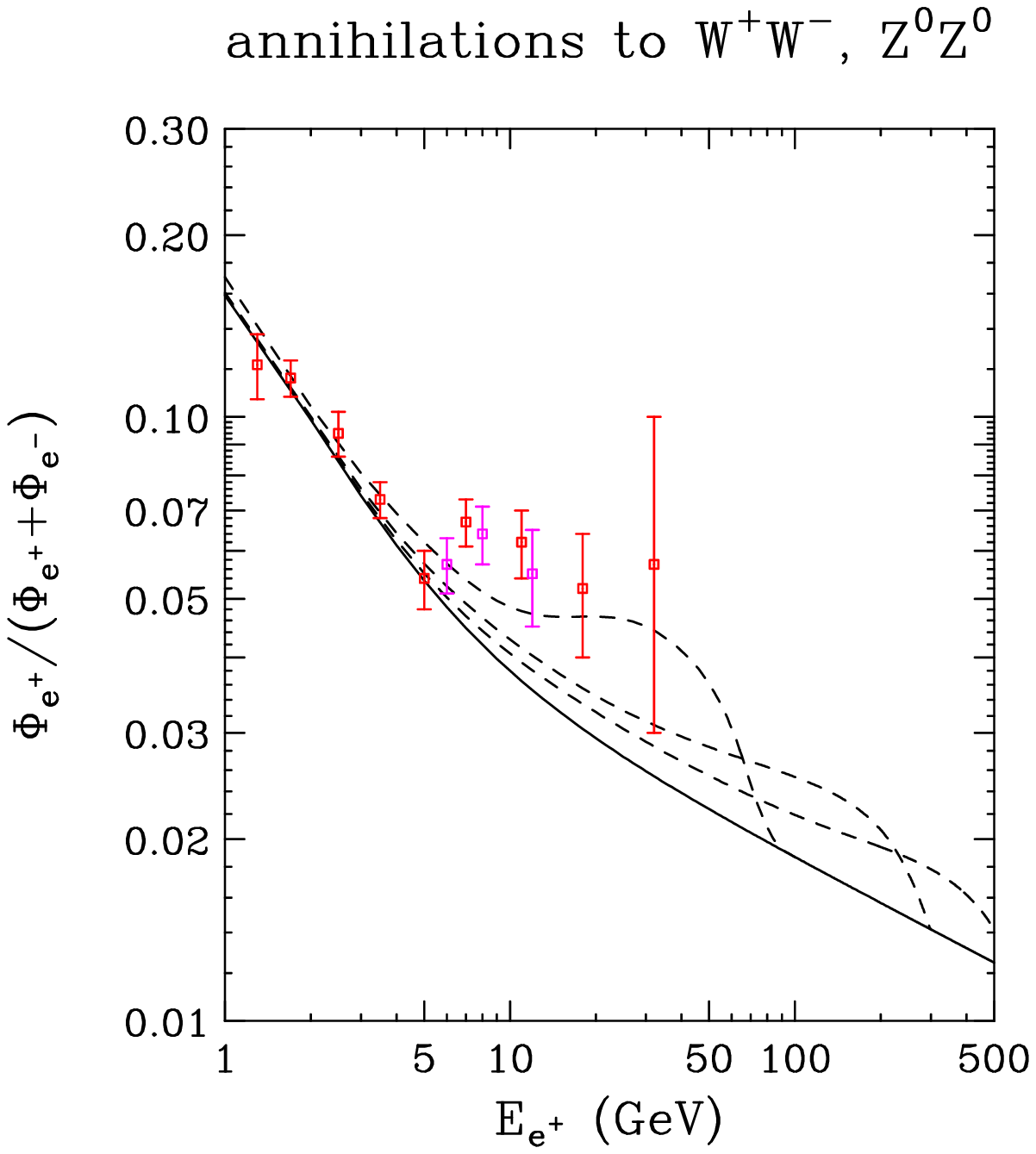}}
\caption{The positron fraction in the cosmic ray spectrum from dark matter annihilations to $ZZ$ and $W^+ W^-$ pairs. WIMP masses of 100, 300 and 600 GeV were considered. A dark matter distribution with $BF=5$ (see section~\ref{cps}), $\rho(\rm{local})=0.43 \,\rm{GeV/cm^3}$ and an annihilation cross section of $\sigma v = 10^{-25} \, \rm{cm}^3/\rm{s}$ was used.}
\label{posgauge}
\end{figure}

\begin{figure}[thb]
\vbox{\kern2.4in\includegraphics{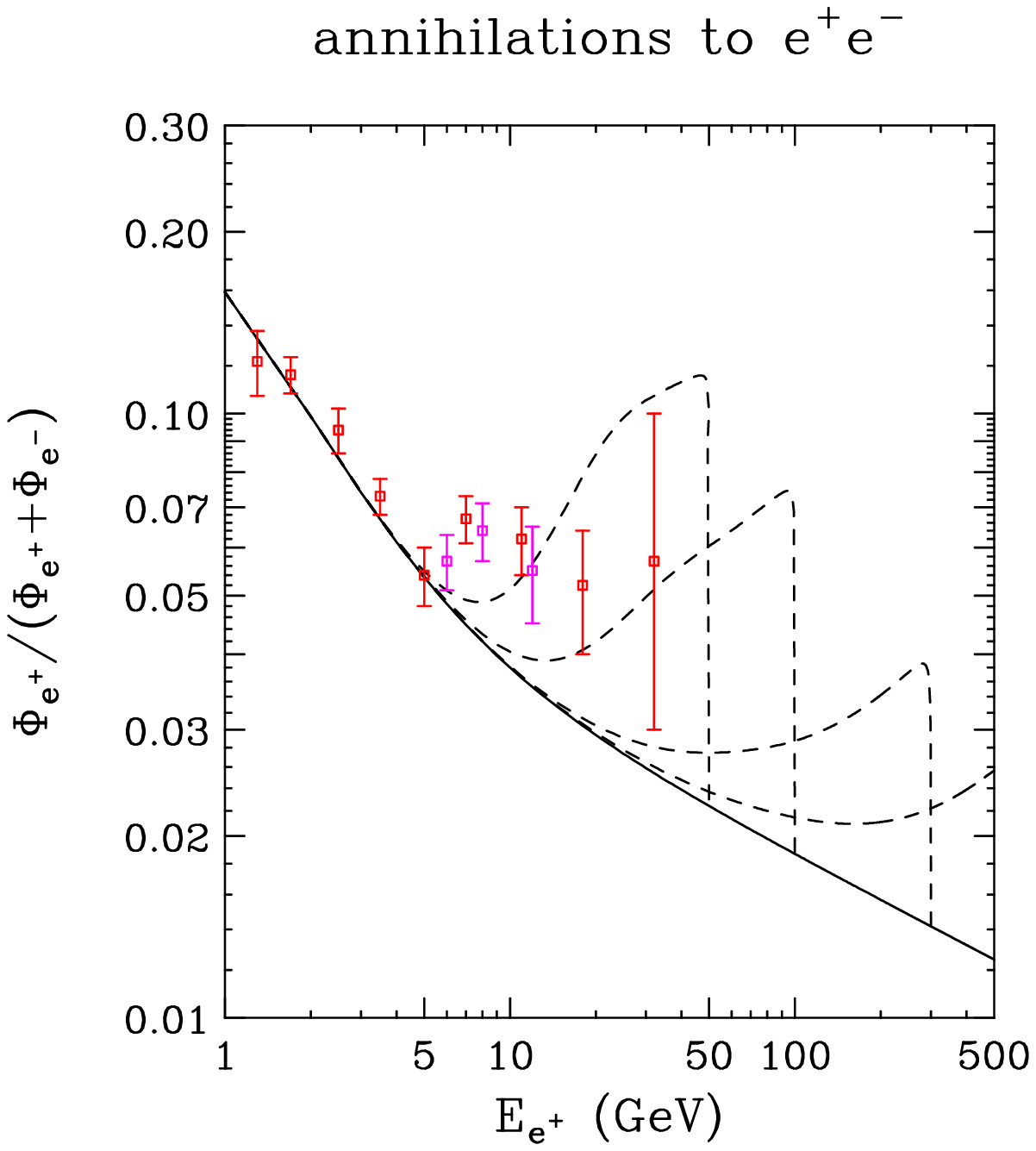}}
\caption{The positron fraction in the cosmic ray spectrum from dark matter annihilations to $e^+ e^-$ pairs. WIMP masses of 50, 100, 300 and 600 GeV were considered. A dark matter distribution with $BF=5$ (see section~\ref{cps}), $\rho(\rm{local})=0.43 \,\rm{GeV/cm^3}$ and an annihilation cross section of $\sigma v = 10^{-26} \, \rm{cm}^3/\rm{s}$ was used.}
\label{posee}
\end{figure}

\begin{figure}[thb]
\vbox{\kern2.4in\includegraphics{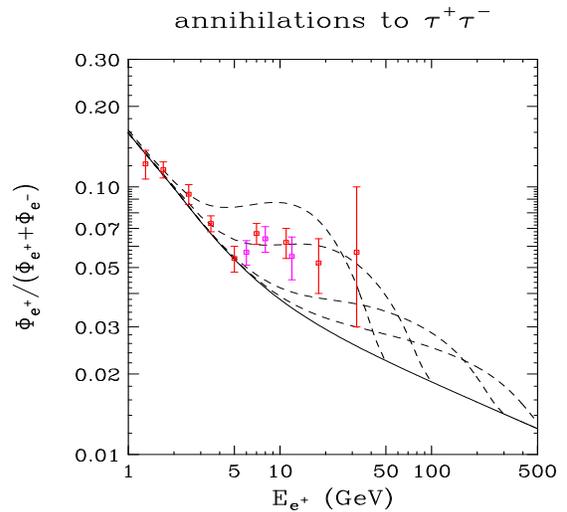}}
\caption{The positron fraction in the cosmic ray spectrum from dark matter annihilations to $\tau^+ \tau^-$ pairs. WIMP masses of 50, 100, 300 and 600 GeV were considered. A dark matter distribution with $BF=5$ (see section~\ref{cps}), $\rho(\rm{local})=0.43 \,\rm{GeV/cm^3}$ and an annihilation cross section of $\sigma v = 10^{-25} \, \rm{cm}^3/\rm{s}$ was used.}
\label{postautau}
\end{figure}

In these figures, the effects of solar modulation are not included. As discussed in the previous section, we have chosen to avoid the uncertainties associated with this effect by studying the ratio of positrons to positrons plus electrons rather than the positron spectrum alone. In figures~\ref{posbb}-\ref{postautau}, the positron fraction is shown for the same annihilation channels. To calculate this quantity, the flux of primary and secondary electrons as well as secondary positrons must be known. We have used the primary and secondary fluxes found in Ref.~\cite{secbg}. Also, in figures~\ref{posbb}-\ref{postautau}, the positron data from the HEAT experiments is shown for comparison.

The magnitude, or normalization, of the positron fluxes shown in figures~\ref{fluxbb}-\ref{postautau} are set by the annihilation rate of dark matter particles. In addition to scaling with the WIMP's annihilation cross section and the square of the local dark matter density, the annihilation rate depends on the degree of inhomogeneity in the local dark matter distribution. This effect is described by a parameter called the {\it boost factor}, $BF$, which is defined by
\begin{equation}
BF=\frac{\int \rho^2 dV}{[\int \rho \, dV]^2},
\end{equation}
where the integral is performed roughly over the volume which contributes substantially to the positron flux (a few kiloparsecs). If the dark matter were locally distributed completely evenly, the boost factor would be equal to one. Small scale clustering of dark matter, however, enhances this quantity to a value expected to be in the range of roughly 2 to 5. Values much larger than this require large local clumps of dark matter and are unlikely to be present \cite{hooperpos}.

\section{Possible Variations}

The results shown in the previous section may be modified with the choice of propagation parameters and Galactic dark matter distribution. In this section, we describe the effect of modifying these choices on the positron spectrum.

\begin{figure}[thb]
\vbox{\kern2.4in\includegraphics{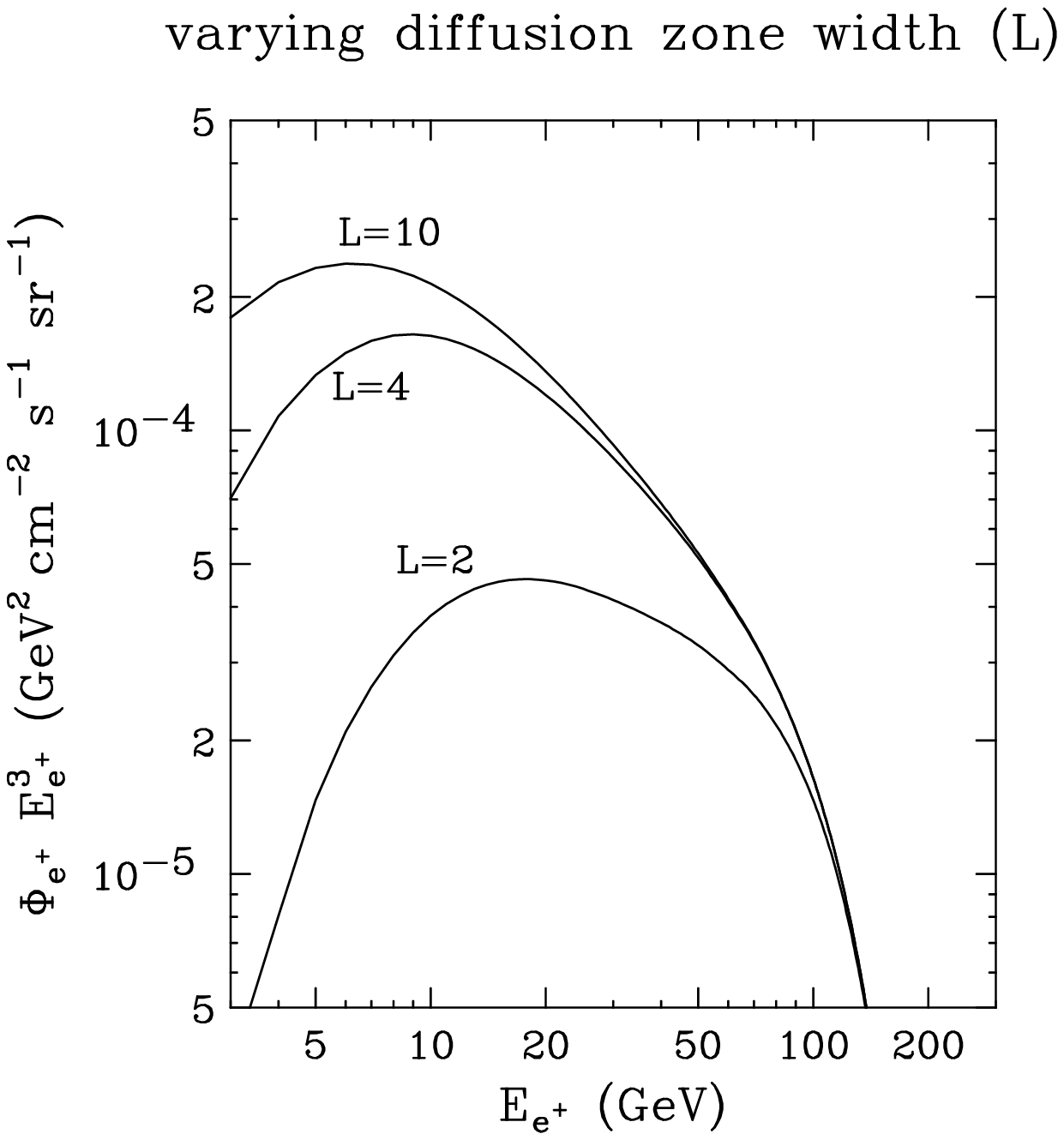}}
\caption{The effect of varying the diffusion zone width ($2L$) on the positron spectrum. Results for $L=2,4$ and 10 kpc are shown. In each case, WIMP annihilations are to b quark pairs and the WIMP's mass is 300 GeV. Other parameters are the same as in the previous figures.}
\label{L}
\end{figure}

\begin{figure}[thb]
\vbox{\kern2.4in\includegraphics{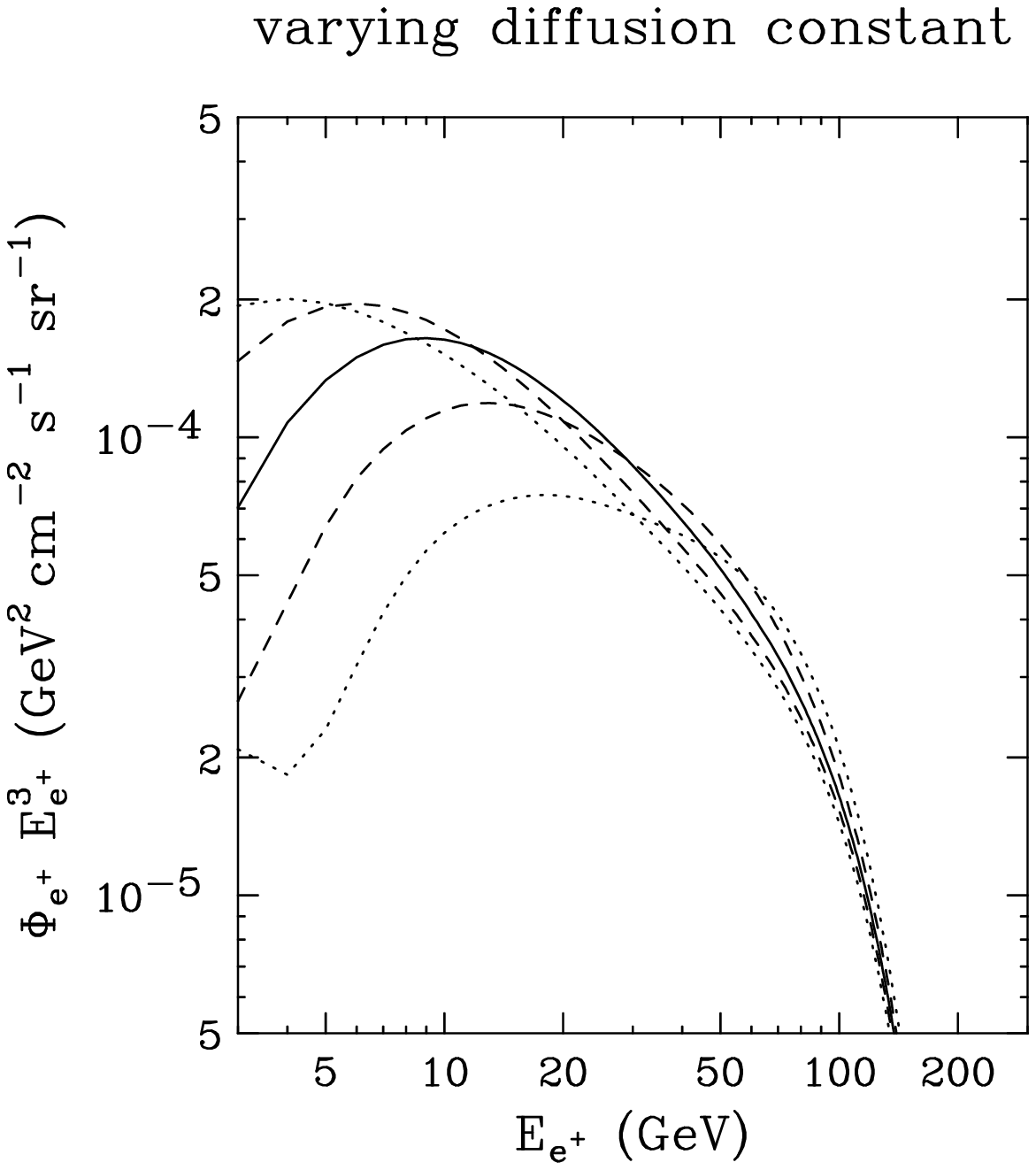}}
\caption{The effect of varying the diffusion constant on the positron spectrum. From top to bottom on the left side of the figure, results are shown for a diffusion constant 0.25 (dotted), 0.5 (dashed), 1.0 (solid), 2.0 (dashed) and 4.0 (dotted) times the value shown in Eq.~\ref{k}. In each case, WIMP annihilations are to b quark pairs and the WIMP's mass is 300 GeV. Other parameters are the same as in the previous figures.}
\label{K}
\end{figure}

\begin{figure}[thb]
\vbox{\kern2.4in\includegraphics{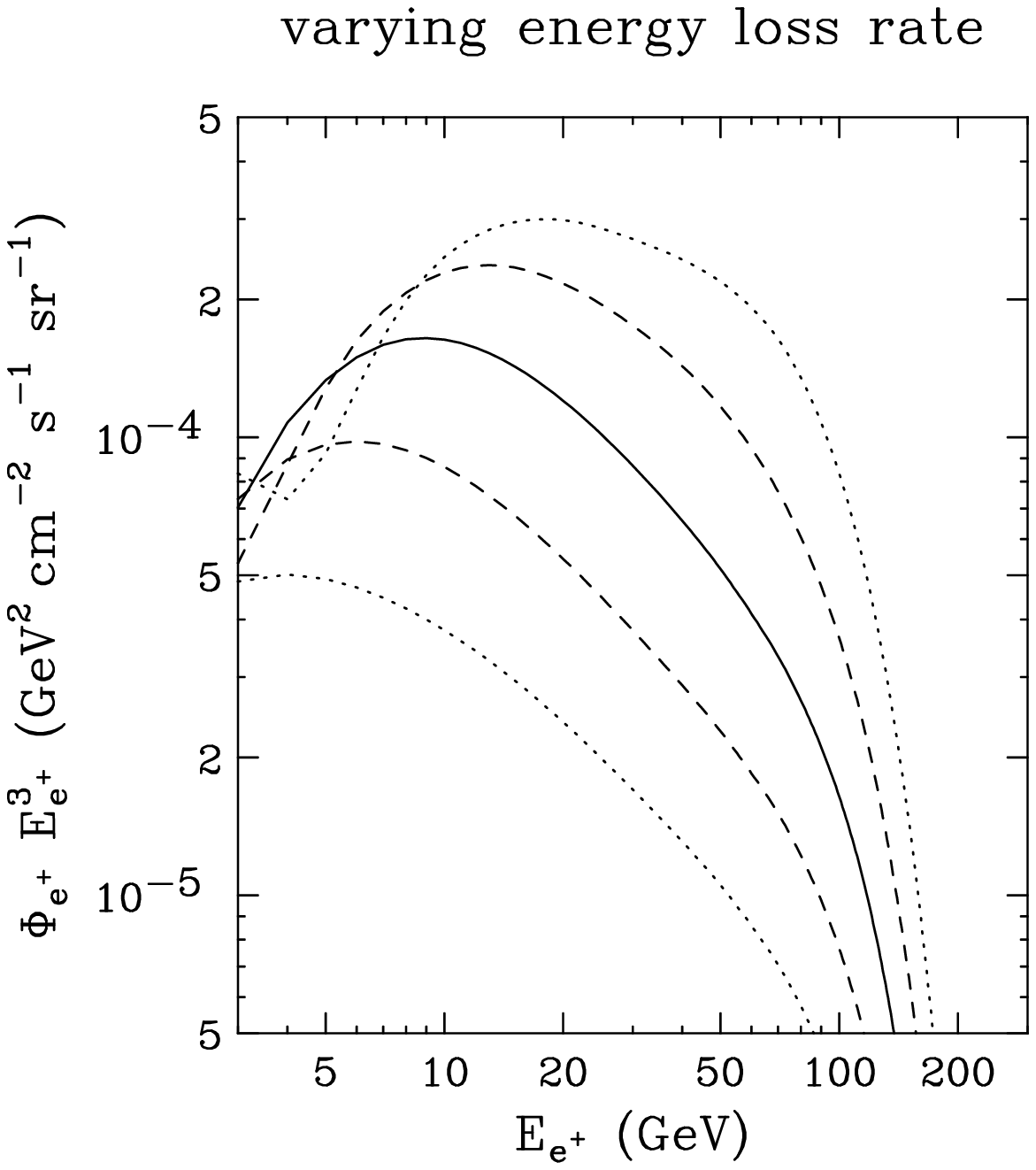}}
\caption{The effect of varying the energy loss rate on the positron spectrum. From top to bottom on the right side of the figure, results are shown for a rate 0.25 (dotted), 0.5 (dashed), 1.0 (solid), 2.0 (dashed) and 4.0 (dotted) times the value shown in Eq.~\ref{b}. In each case, WIMP annihilations are to b quark pairs and the WIMP's mass is 300 GeV. Other parameters are the same as in the previous figures.}
\label{bvary}
\end{figure}

\begin{figure}[thb]
\vbox{\kern2.4in\includegraphics{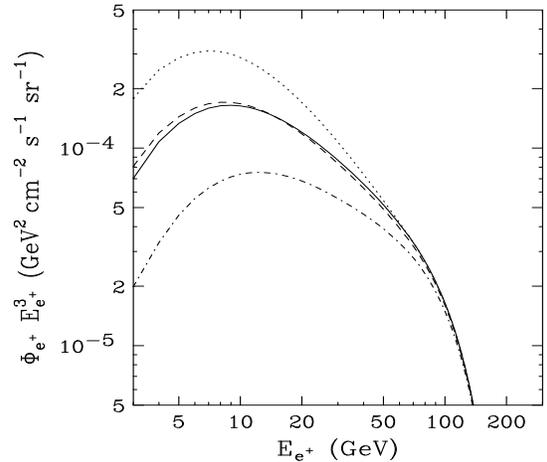}}
\caption{The effect of varying the dark matter halo profile on the positron spectrum. From top to bottom on the left side of the figure, results are shown for a Moore {\it et al.} profile (dotted), an Navarro-Frenk-White profile (dashed), and isothermal sphere profile (solid) and a completely flat distribution (dot-dashed). In each case, WIMP annihilations are to b quark pairs and the WIMP's mass is 300 GeV. Other parameters are the same as in the previous figures.}
\label{profile}
\end{figure}

The diffusion parameters which must be set in our formalism are the diffusion zone width, $2L$, the energy loss rate, $b$, and the diffusion constant, $K$, including its energy dependence. The impact of varying these parameters is shown in figures~\ref{L}-\ref{profile}.

The effect of modifying the diffusion zone width is shown in figure~\ref{L}.  The flux at high energies is less affected by the choice of $L$ than at lower energies. This is because distant annihilations produce positrons which are substantially degraded in energy before they reach Earth. Thus nearby annihilations produce most of the positrons at the high end of the spectrum. As the width of the diffusion zone is reduced, fewer positrons reach Earth at lower energies.

Variations in the magnitude of the diffusion constant of Eq.~\ref{k} can also have an impact on the positron flux observed. As this quantity is increased, the lowest energy positrons are most dramatically effected. Positrons of higher energies are actually mildly enhanced as the diffusion constant is increased.

\newpage

\vspace{1.0cm}

\vspace{1.0cm}

\vspace{1.0cm}

\newpage

\vspace{1.0cm}

\vspace{1.0cm}

\vspace{1.0cm}

\newpage

Modifications of the positron energy loss rate (or equivalently the density of diffuse starlight and cosmic microwave background photons) is tied to changes in the diffusion constant. Inspection of Eq.~\ref{dif} reveals that the resulting spectral shape (up to normalization) is invariant with respect to the quantity $K(E_{e^{+}},\vec{x})/b(E_{e^{+}},\vec{x})$. The normalization of the positron flux, on the other hand, scales inversely with either the diffusion constant or the energy loss rate. 

Finally, the choice of dark matter halo profile model considered can affect the positron flux observed. Thus far, our results have used an isothermal sphere dark matter profile, described by
\begin{equation}
\rho(r) = 0.43 \, \rm{GeV/cm^3} \, \frac{(2.8 \, \rm{kpc})^2 + (8.5 \, \rm{kpc})^2} {(2.8 \, \rm{kpc})^2 + r^2},
\end{equation}
where $r$ is the distance from the Galactic Center. This profile exhibits a fairly flat behavior in the inner few kiloparsecs of the Galaxy. Alternatively, we may consider a dark matter distribution favored by N-body simulations, such as the Navarro-Frenk-White (NFW) \cite{nfw} profile:
\begin{equation}
\rho(r) = 0.43 \, \rm{GeV/cm^3} \, \frac{(8.5/20) \, [1+(8.5/20)]^2} {(r/20 \,\rm{kpc}) \, [1+(r/20 \,\rm{kpc})]^2},
\end{equation}
or the Moore {\it et al.} profile \cite{moore}:
\begin{equation}
\rho(r) = 0.43 \, \rm{GeV/cm^3} \, \frac{(8.5/28)^{1.5} \, [1+(8.5/28)^{1.5}]} {(r/28 \,\rm{kpc})^{1.5} \, [1+(r/28 \,\rm{kpc})^{1.5}]}.
\end{equation}
Unlike the isothermal sphere profile, the NFW and Moore {\it et al.} profiles have density cusps at the Galactic Center. Each of these profiles have been normalized to a local dark matter density of $0.43 \rm{GeV/cm^3}$. The only locations which can deviate dramatically from this value are in the region near the Galactic Center. Sharply cusped profiles, the Moore {\it et al.} profile in particular, produce more positrons in the inner Galaxy and therefore more lower energy positrons are observed locally. This effect is shown in figure~\ref{profile}. The NFW and isothermal profiles produce very similar results, with fewer low energy positrons than the Moore {\it et al}. case. Shown for comparison is the result for a completely flat dark matter distribution, $\rho(r) =\, \rm{constant}\, = 0.43 \,\rm{GeV/cm^3}$.

\section{Neutralino Dark Matter}
\label{neudm}

In many models of supersymmetry, the Lightest Supersymmetric Particle (LSP) is a neutralino \cite{susylsp}, a mixture of the superpartners of the photon, $Z$ and neutral Higgs bosons. In models of supersymmetry with conserved R-parity, the LSP is stable and a potentially viable dark matter candidate. In this section, we will discuss the signatures of neutralino annihilations in the cosmic positron spectrum.

The dominant annihilation channels of a neutralino depend on its composition. If the neutralino is mostly bino-like (the superpartner of the hypercharge gauge boson), it will annihilate largely to heavy fermions, {\it i.e.} $b \bar{b},\, \tau^+ \tau^- \,$and, if kinematically allowed, $t \bar{t}$. Annihilations to lighter fermions are rare as the low velocity neutralino annihilation cross section is chirality suppressed by a factor of $m_{\rm{f}}^2/m_{\chi^0}^2$. Furthermore, the annihilation cross section to down-type fermions, such as $b \bar{b}$, is enhanced in some diagrams by a factor of $\tan^2 \beta$. In much of the supersymmetric parameter space, the LSP annihilates almost entirely to $b \bar{b}$ with a small admixture of $\tau^- \tau^+$. If $\tan \beta$ is fairly small and the LSP mass is greater than the top mass, annihilations to $t \bar{t}$ can also be important.

The lightest neutralino might not be mostly bino, however. Alternatively, it may be largely higgsino-like (superpartner of the neutral Higgs bosons). In this case, the LSP may annihilate primarily to gauge boson pairs. The low velocity annihilation cross sections for a pure higgsino are approximately given by:
\begin{equation}
<\sigma v>_{W^+ W^-} \simeq \bigg(1- \frac{m_W^2}{m_{\chi^0}^2}  \bigg) \frac{G_F^2 m_W^4 m_{\chi^0}^2}{\pi (2 m_{\chi^0}^2-m_W^2)^2}
\end{equation}
and
\begin{equation}
<\sigma v>_{Z^0 Z^0} \simeq \bigg(1- \frac{m_Z^2}{m_{\chi^0}^2}  \bigg) \frac{G_F^2 m_Z^4 m_{\chi^0}^2}{2\pi (2 m_{\chi^0}^2-m_Z^2)^2}.
\end{equation}
These cross sections can be quite large, reaching $3 \times 10^{-25}\,\rm{cm^3}/\rm{s}$ at their maximum near $m_{\chi^0}=110$ GeV. A mixed higgsino-bino LSP might also annihilate mostly to gauge bosons, but with a smaller cross section.

As a third possibility, the lightest neutralino may also be a wino (the superpartner of the isospin gauge boson). Although wino-like LSPs are uncommon in most supersymmetry breaking scenarios, in the interesting case of Anomaly Mediated Supersymmetry Breaking (AMSB), the lightest neutralino is a nearly pure wino. This is because the ratios of the gaugino masses do not follow the common hierarchy ($M_1 < M_2 < M_3$), but rather are proportional to $\beta$-functions resulting in the ratios $M_1 : M_2 : M_3 = 2.8 : 1 : 7.1$. When this ratio of $M_1$ to $M_2$ is inserted into the neutralino mass matrix, some generic phenomenological features of this model emerge. In addition to the LSP being a neutral wino, a chargino only a few hundred MeV heavier than the LSP is present. This can lead to efficient coannihilations in the relic density calculation. Much like the higgsino case, wino-like neutralinos in AMSB annihilate primarily via gauge bosons channels \cite{amsb}.

\newpage

\vspace{1.0cm}

\vspace{1.0cm}

\vspace{1.0cm}

\newpage

\begin{figure}[thb]
\vbox{\kern2.4in\includegraphics{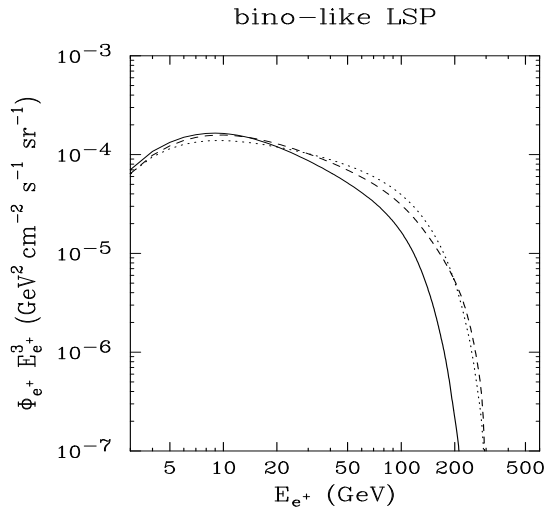}}
\caption{The spectrum of positrons, including the effects of propagation, from bino-like neutralino annihilations. The solid line is for annihilations purely to $b \bar{b}$. The dashed line is for annihilations to $b \bar{b}$ with a 10\% admixture to $\tau^+ \tau^-$. The dotted line represents 50\% to $b \bar{b}$ and 50\% to  $t \bar{t}$. For each case a 300 GeV neutralino was considered. A dark matter distribution with $BF=5$ (see section~\ref{cps}), $\rho(\rm{local})=0.43 \,\rm{GeV/cm^3}$ and an annihilation cross section of $\sigma v = 10^{-25} \, \rm{cm}^3/\rm{s}$ was used. The effects of solar modulation are not included.}
\label{fluxbino}
\end{figure}

We now turn our attention to these various types of neutralino LSPs in the context of their annihilations producing positrons. The spectral shape of the positrons produced in the annihilation of neutralino LSPs is often dominated by either $b \bar{b}$ or gauge boson channels. In the bino case, as we said before, annihilations are typically almost entirely to $b \bar{b}$ with only a few percent admixture to $\tau^+ \tau^-$. To see the effect of this admixture, we compare the positron spectrum from annihilations purely to $b \bar{b}$ with the spectrum from annihilations 90\% to $b \bar{b}$ and 10\% to $\tau^+ \tau^-$. Below about $E_{e^+} \sim m_{\chi^0}/10$, the spectrum is not significantly effected by the tau component.  At higher energies the tau channel becomes more important. For a 10\% $\tau^+ \tau^-$ admixture, the flux at $E_{e^+} \sim m_{\chi^0}/2$ is roughly doubled over the pure $b \bar{b}$ case. For an equal fraction of annihilations into $b \bar{b}$ and $t \bar{t}$, the result is similar. These spectra are shown in figure~\ref{fluxbino}. We have already shown the positron spectra for annihilations to gauge boson pairs, which are expected for higgsino-like or wino-like neutralinos.

\vspace{1.0cm}

Predicting the spectral shape of the positrons produced in dark matter annihilations is only useful if the flux is large enough to be observed over background. The most important quantity in determining this flux is the dark matter annihilation rate, which in turn depends on the low velocity WIMP annihilation cross section. 

If the dark matter is made up thermal relics produced in the early universe, its density today can be calculated in a straight forward way. The result of this standard technique \cite{review} is 
\begin{equation}
\Omega_{\rm{WIMP}} h^2 \cong \frac{1.07 \times 10^9 \, \rm{GeV}^{-1}}{M_{\rm{Pl}}} \frac{x_F}{\sqrt{g_{\star}}} \frac{1}{(a+3b/x_F)},
\label{relic1}
\end{equation}
where $M_{\rm{Pl}}$ is the Plank mass, $g_{\star}$ is the number of relativistic degrees of freedom assessable at freeze-out and $a$ and $b$ are the first and second terms in the expansion of the annihilation cross section: $<\sigma v> = a +b v^2 + \mathcal{O}$$(v^4)$. $x_F$ is the WIMP mass over the temperature at freeze-out, and is generally about 20 for a weakly interacting dark matter candidate. This expression neglects the possible role of coannihilations between the WIMP and other particles during freeze-out. Coannihilations may be important when the LSP is only slightly lighter than the Next-to-Lightest Supersymmetric Particle (NLSP). In such a case, the neutralino relic density can be depleted below the value of Eq.~\ref{relic1} and therefore a smaller annihilation cross section would be required to provide the measured relic abundance.

\begin{figure}[thb]
\vbox{\kern2.4in\includegraphics{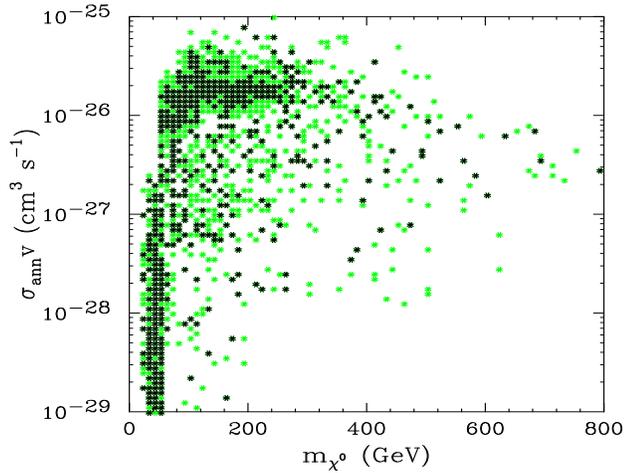}}
\caption{The neutralino annihilation cross section in the low velocity limit for parameters within the MSSM which yield a thermal relic density in the range measured by WMAP ($\Omega_{\rm{CDM}} h^2 = 0.129-0.095$, dark shading) or a somewhat larger range ($\Omega_{\rm{CDM}} h^2 = 0.16-0.06$, light green shading). For all models shown, the ratio of Higgs expectation values, $\tan \beta$, is set to 5.}
\label{sigma5}
\end{figure}

\begin{figure}[thb]
\vbox{\kern2.4in\includegraphics{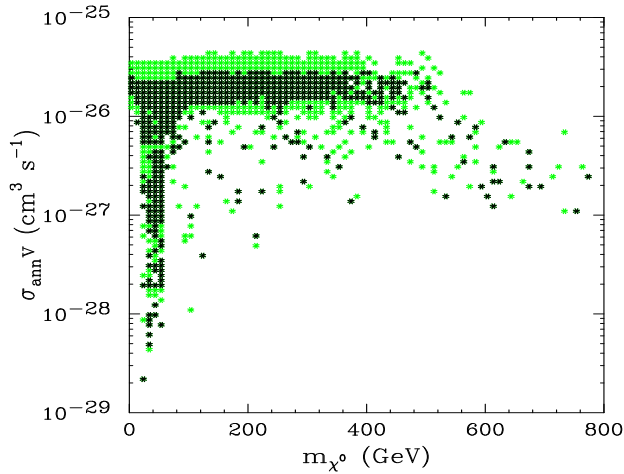}}
\caption{The neutralino annihilation cross section in the low velocity limit for parameters within the MSSM which yield a thermal relic density in the range measured by WMAP ($\Omega_{\rm{CDM}} h^2 = 0.129-0.095$, dark shading) or a somewhat larger range ($\Omega_{\rm{CDM}} h^2 = 0.16-0.06$, light green shading). For all models shown, the ratio of Higgs expectation values, $\tan \beta$, is set to 50.}
\label{sigma50}
\end{figure}

For a weakly interacting dark matter candidate, Eq.~\ref{relic1}  numerically yields: 
\begin{equation}
\Omega_{\rm{WIMP}} h^2 \sim  \frac{3 \times 10^{-27}\,\rm{cm^3}/\rm{s}}{(a+3b/20)}.
\end{equation}
Matching this to the cold dark matter density observed by WMAP, $\Omega_{\rm{WIMP}} h^2 \simeq 0.1$, we arrive at
\begin{equation}
a+3b/20 \sim   3 \times 10^{-26}\,\rm{cm^3}/\rm{s}.
\label{abeq}
\end{equation}
The low velocity annihilation cross section, $a$, for neutralino LSPs with a thermal relic density in the range of $0.095 < \Omega_{\chi^0} h^2 < 0.129$ (dark shading) and $0.06 < \Omega_{\chi^0} h^2 < 0.16$ (light green shading) is shown in figures~\ref{sigma5} and~\ref{sigma50} for models with $\tan \beta$ equal to 5 and 50, respectively. In the high $\tan \beta$ case shown in figure~\ref{sigma50}, low velocity annihilations to $b \bar{b}$ are highly efficient and, for neutralinos heavier than about 80 GeV, the low velocity cross section is often roughly the maximum value allowed by Eq.~\ref{abeq}, $<\sigma v > \sim a \sim  3 \times 10^{-26}\,\rm{cm^3}/\rm{s}$. For smaller values of $\tan \beta$, the $b$-term in the velocity expansion of the annihilation cross section can be important, thus allowing a wide range of low velocity cross sections as shown in figure~\ref{sigma5}.

The points shown in figures~\ref{sigma5} and~\ref{sigma50} were calculated using a variation of the DarkSusy package \cite{darksusy}, varying the parameters $\mu$, $M_1$, $M_2$, $M_3$, $A_t$, $A_b$ and the sfermion masses randomly up to 10,000 GeV, and with either sign when appropriate. The pseudoscalar Higgs mass was varied up to 1 TeV. Although the complete MSSM contains more than 100 free parameters, the results shown in figures~\ref{sigma5} and~\ref{sigma50} represent an illustrative sample of the possible models. Note that the sampling technique used in our monte carlo is somewhat limited for finding models with an LSP heavier than several hundred GeV. Such models certainly exist, although they are probably also quite fine tuned.

\vspace{1.0cm}

\vspace{1.0cm}

From the perspective of the calculation of the thermal relic density, higgsino or wino-like LSPs may appear disfavored, as their annihilation cross section is often too large to satisfy conditions such as those in Eq.~\ref{abeq}. In fact, to provide the thermal relic density observed by WMAP a pure higgsino must have a mass of about 1 TeV. A pure-wino would need to be even more massive to provide the observed quantity of dark matter thermally. 

Alternatively, neutralino LSPs may be produced by non-thermal mechanisms thereby evading the requirements of 	Eq.~\ref{abeq}. In some scenarios, objects with small annihilation cross sections may freeze-out only to decay into LSPs at a later time. Long lived, but unstable states, such as gravitinos or Q-balls, would be examples of such intermediate states. In a non-thermal mechanism such as this, the LSP's annihilation cross section can be considerably larger than in the standard thermal relic case \cite{nonthermal}.

For a more detailed discussion of neutralino dark matter in various supersymmtry breaking scenarios, see Ref.~\cite{hooperwang}.

\section{Kaluza-Klein Dark Matter}
\label{kkdmsec}

In models with extra spatial dimensions, Kaluza-Klein (KK) excitations of Standard Model particles may appear, and in some models may provide a viable dark matter candidate \cite{kkdark,taitservant}. In particular, in models with Universal Extra Dimensions (UED), in which all of the fields of the Standard Model are free to propagate in the bulk, the Lightest Kaluza-Klein Particle (LKP) may be stable \cite{universal}. This is because the conservation of momentum in higher dimensional space is conserved, and therefore the KK number of a state is conserved. For a UED model to be phenomenologically viable, the extra compact dimensions must be modded out by an orbifold. Orbifolding leads to terms which violate KK number, but may leave a conserved quantity, called KK-parity, which ensures that the LKP is stable in much the same that R-parity stabilizes the LSP in supersymmetry.

At tree level, all of the first level KK modes have a mass corresponding to the compactifaction scale of the extra dimensions and their zero-mode mass: $(m^(1)_i)^2 = (m^(0)_i)^2 +  1/R^2 $. Radiative corrections to the masses of the tree level spectrum break the (near) degeneracy of the first level KK states \cite{radiative}. The lightest KK state is most naturally the level one excitation of the hypercharge gauge boson, $B^(1)$. For the remainder of this article, we will refer to this state simply as the LKP or Kaluza-Klein Dark Matter (KKDM). 

Although this LKP is similar to a bino-like neutralino in many phenomenological respects, its characteristics differ in significant ways. Perhaps most importantly for our purposes, the LKP is a boson and therefore its annihilations to fermions are not chirality suppressed. Thus, unlike neutralino dark matter, KKDM can annihilate directly to $e^+e^-$, $\mu^+ \mu^-$ and $\tau^+ \tau^-$, which each yield a generous number of high energy positrons. The annihilation cross section of KKDM is nearly proportional to the hypercharge of the final state fermions to the fourth power, therefore its annihilations are primarily to charged leptons (approximately $20\%$ per generation). The remaining annihilations are to up-type quarks (approximately $11\%$ per generation), neutrinos (approximately $1.2\%$ per generation), Higgs bosons (approximately $2.3\%$) and down-type quarks  (approximately $0.7\%$ per generation). 

The total annihilation cross section of KKDM is
\begin{equation}
<\sigma v> = \frac{95 g_1^4}{324 \pi m^2_{LKP}} \simeq \frac{1.7 \times 10^{-26} \, \rm{cm^3}/\rm{s}}{m^2_{LKP}(\rm{TeV})}. 
\end{equation}
This annihilation cross section consists entirely of an $a$-term in the expansion $<\sigma v> = a + b v^2 + \mathcal{O}$$(v^4)$, {\it i.e.} $b=0$ and $<\sigma v> \cong a$. Comparing this expression with Eq.~\ref{abeq}, we find that the thermal relic density of KKDM matches the value observed by WMAP for a mass of about $m_{LKP} \approx 800$ GeV, neglecting the effect of coannihilations \cite{taitservant}. Coannihilations can play an important role in the freeze-out of KKDM, however. It is somewhat expected that the loop-level radiative corrections to the KK spectrum should be much smaller than the tree-level contribution, thus there should be many KK states only slightly more massive than the LKP. Take, for example, one of the cases studied by Tait and Servant \cite{taitservant}. In this scenario, the first level KK excitations of $e_R$ were nearly degenerate with the LKP. The $e_R^{(1)}-e_R^{(1)}$ annihilation cross section is only slightly larger than the LKP-LKP annihilation cross section and the $e_R^{(1)}$-LKP coannihilation cross section is substantially smaller. Thus each species freeze-out quasi-independently and after freeze-out, the remaining $e_R^{(1)}$'s decay to LKPs thus {\it enhancing} the LKP relic density. This is in contrast to the coannihilation mechanism often considered for neutralinos, in which coannihilations between the LSP and other supersymmetric particles can be large and efficiently {\it deplete} the neutralino relic density. 

Tait and Servant find that for three generations of $e_R^{(1)}$'s, each 1\% heavier that the LKP, a mass of $m_{LKP} \cong 550-650$ GeV is needed to produce the observed density of LKP dark matter \cite{taitservant}. Furthermore, other coannihilation channels may certainly play an important role in determining the relic density of KKDM, plausibly lowering the mass of the LKP considerably. Regardless of the coannihilation channels available, the LKP mass is constrained by electroweak precision measurements to be heavier than about 300 GeV \cite{ew300}.

\newpage

\vspace{1.0cm}

\begin{figure}[thb]
\vbox{\kern2.4in\includegraphics{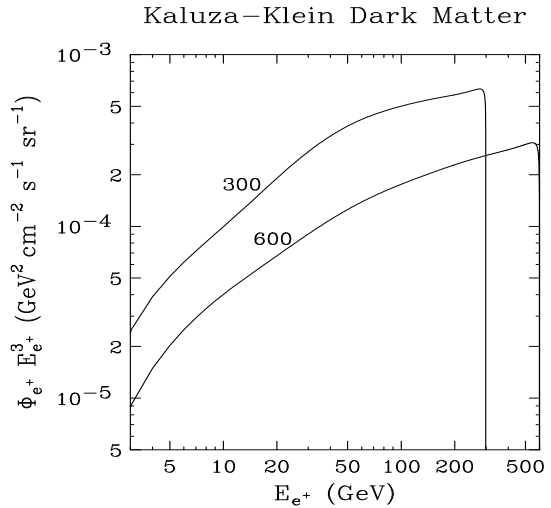}}
\caption{The spectrum of positrons including the effects of propagation from Kaluza-Klein Dark Matter (KKDM) annihilations. Annihilations of KKDM produce equal fractions of $\tau^+ \tau^-$, $\mu^+ \mu^-$ and $e^+ e^-$ pairs (approximately $20\%$ each) as well as up-type quarks (approximately $11\%$ per generation), neutrinos (approximately $1.2\%$ per generation), Higgs bosons (approximately $2.3\%$) and down-type quarks  (approximately $0.7\%$ per generation). Results for KKDM masses of 300 and 600 GeV are shown. A dark matter distribution with $BF=5$ (see section~\ref{cps}), $\rho(\rm{local})=0.43 \,\rm{GeV/cm^3}$ and an annihilation cross section of $\sigma v = 10^{-25} \, \rm{cm}^3/\rm{s}$ was used. The effects of solar modulation are not included.}
\label{fluxKK}
\end{figure}

\vspace{1.0cm}

\begin{figure}[thb]
\vbox{\kern2.4in\includegraphics{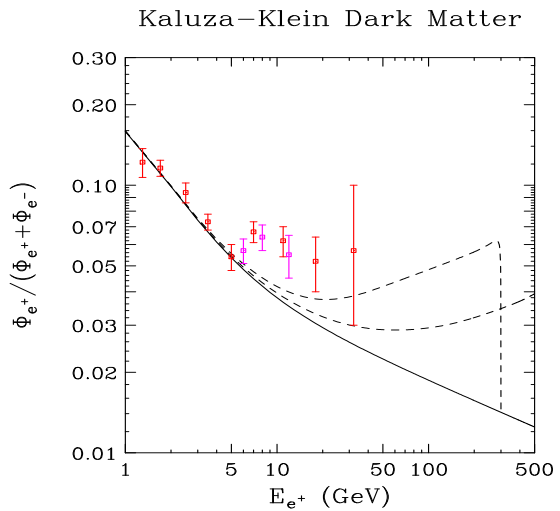}}
\caption{The positron fraction in the cosmic ray spectrum from Kaluza-Klein Dark Matter (KKDM) annihilations. KKDM masses of 300 and 600 GeV were considered.  A dark matter distribution with $BF=5$ (see section~\ref{cps}), $\rho(\rm{local})=0.43 \,\rm{GeV/cm^3}$ and an annihilation cross section of $\sigma v = 10^{-25} \, \rm{cm}^3/\rm{s}$ was used.}
\label{posKK}
\end{figure}

\newpage

\section{The HEAT Results}

The HEAT (High-Energy Antimatter Telescopes) experiment's balloon flights in 1994-95 and 2000 have measured the cosmic positron spectrum between energies of approximately 1 to 30 GeV \cite{heat1995,heat2000,heat}. When this data is presented as a positron fraction, an interesting feature appears. The origin of this feature, a bump appearing at 7-10 GeV, is not understood. Although many efforts have been made to fit this feature with a contribution from annihilating dark matter, it has been shown to be very difficult to satisfactorily match the spectrum of the bump \cite{positrons3,posbaltz,kkpos,maor}.

Far more statistically significant than the bump-feature present in the HEAT data, however, is the overall excess of positrons above the background prediction at energies above about 7 GeV. This, in principle, can certainly be accommodated by a contribution to the positron spectrum from dark matter annihilations, although rather large cross sections or boost factors are required to do so.

To judge the degree that adding a dark matter component to the positron spectrum improves the fit to the HEAT data, we calculate the $\chi^2$,
\begin{equation}
\chi^2 = \sum \frac{(N_{\rm{Obs}}-N_{\rm{DM}}-N_{\rm{BG}})^2}{\Delta_{N_{\rm{Obs}}}},
\label{chi2}
\end{equation}
where the sum is over energy bins (one for each HEAT error bar), $N_{\rm{Obs}}$ is the number of events observed in that bin, $N_{\rm{DM}}$ is the number of events predicted from the annihilating dark matter contribution, $N_{\rm{BG}}$ is the number of events predicted from the background contribution and $\Delta_{N_{\rm{Obs}}}$ is the error associated with the measurement in each energy bin. 

The $\chi^2$ we find for the background-only curve is 47.2 over 12 degrees of freedom (error bars). This $\chi^2$ of almost 4 per degree of freedom represents a very poor fit to the HEAT data. The overall quality of this fit can be dramatically improved by including a new component from dark matter annihilations.

\begin{figure}[thb]
\vbox{\kern2.4in\includegraphics{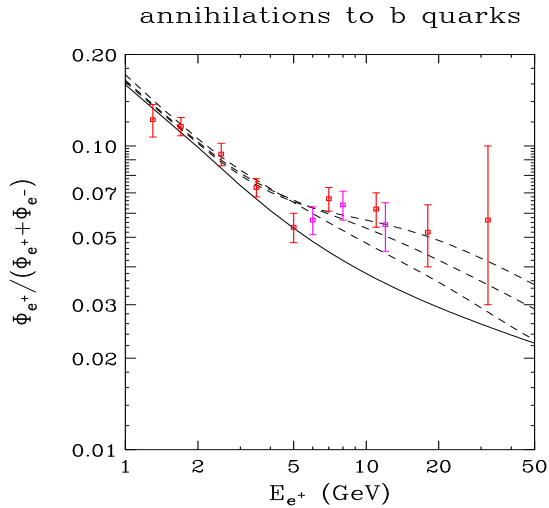}}
\caption{The positron fraction for dark matter annihilations to $b\bar{b}$ for WIMP masses of 100, 300 and 600 GeV (bottom to top on the right side of the figure). In each case, the normalization of the dark matter contribution was chosen to maximize the quality of the fit to the HEAT data (shown as error bars). We find $\chi^2$'s per degree of freedom of 23.5/12, 13.6/12 and 9.7/12 for 100, 300 and 600 GeV WIMPs, respectively. Considering a dark matter distribution with $BF=5$ (see section \ref{cps}) and $\rho$(local)=0.43 GeV/cm$^3$, annihilation cross sections of $\sigma v = 9 \times 10^{-26}\,\rm{cm^3/s}$, $2.5 \times 10^{-25}\,\rm{cm^3/s}$ and $5.5 \times 10^{-25}\,\rm{cm^3/s}$ are needed for these fits for 100, 300 and 600 GeV WIMPs, respectively. The solid line is the background-only prediction.}
\label{heatbb}
\end{figure}

Considering first annihilations primarily to $b \bar{b}$, we find that only rather heavy WIMPs ($m \gsim 200\, $GeV) can accommodate the HEAT data. This can be seen from the 100 GeV WIMP curve shown in figure~\ref{heatbb}. For this curve, the flux at low energies consistently exceeds the HEAT measurements while it falls below the measurements above 10 GeV. We find that for the best-fit normalization ($\sigma v = 9 \times 10^{-26}\,\rm{cm^3/s}$, $BF=5$, $\rho_{\rm{local}}$=0.43 GeV/cm$^3$) a $\chi^2$ of 23.5, or almost 2 per degree of freedom, is obtained, representing a rather poor fit. For a 300 or 600 GeV WIMP annihilating to $b \bar{b}$, however, we find a $\chi^2$ of 13.6 and 9.7, respectively. In these cases, the normalization must be increased significantly, however. Using $BF=5$ and $\rho_{\rm{local}}$=0.43 GeV/cm$^3$, we must require $\sigma v = 2.5 \times 10^{-25}\,\rm{cm^3/s}$ and $\sigma v = 5.5 \times 10^{-25}\,\rm{cm^3/s}$ to achieve these good fits. These cross sections are each well above the value acceptable for a thermal relic (see Eq.~\ref{abeq}).

\vspace{1.0cm}

\begin{figure}[thb]
\vbox{\kern2.4in\includegraphics{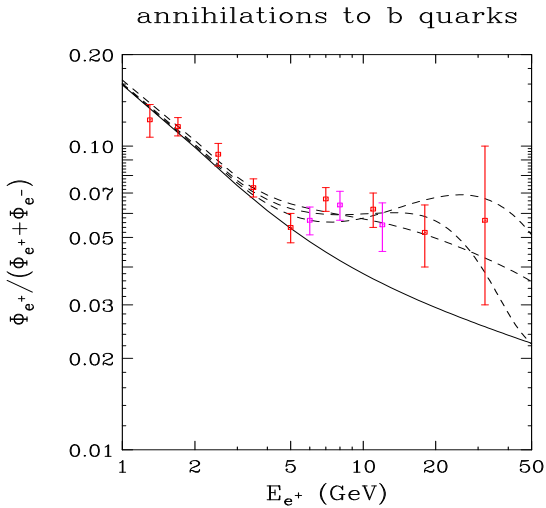}}
\caption{The same as in figure~\ref{heatbb}, except with a diffusion constant twice as large. WIMPs of mass 50, 100 and 300 GeV are shown (100, 300, 50 from top to bottom on the right). Here we find $\chi^2$'s per degree of freedom of 7.5/12, 9.7/12 and 9.5/12 for 50, 100 and 300 GeV masses, respectively. Considering a dark matter distribution with $BF=5$ (see section \ref{cps}) and $\rho$(local)=0.43 GeV/cm$^3$, annihilation cross sections of $\sigma v = 6 \times 10^{-26}\,\rm{cm^3/s}$, $1.5 \times 10^{-25}\,\rm{cm^3/s}$ and $4.5 \times 10^{-25}\,\rm{cm^3/s}$ are needed for these fits for 50, 100 and 300 GeV WIMPs, respectively. The solid line is the background-only prediction.}
\label{heatbbdif}
\end{figure}

It is worth noting that the situation shown in figure~\ref{heatbb} can be improved if more favorable diffusion and halo parameters are used. For example, if we use a diffusion constant twice the size of our standard choice (shown in Eq.~\ref{k}), we can improve the fit to the HEAT data substantially, consistent with WIMPs as light as 30-40 GeV. These results are shown in figure~\ref{heatbbdif}.

\begin{figure}[thb]
\vbox{\kern2.4in\includegraphics{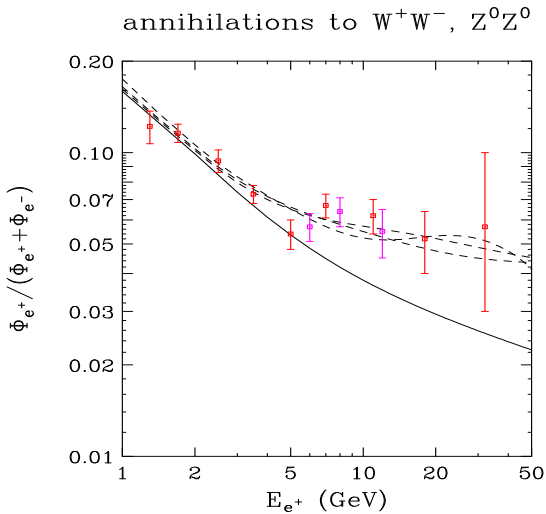}}
\caption{The positron fraction for dark matter annihilations to $ZZ$ and $W^+ W^-$ pairs for WIMP masses of 300, 600 and 100 GeV (bottom to top at 30 GeV). In each case, the normalization of the dark matter contribution was chosen to maximize the quality of the fit to the HEAT data (shown as error bars). We find $\chi^2$'s per degree of freedom of 15.3/12, 11.4/12 and 8.8/12 for 100, 300 and 600 GeV WIMPs, respectively. Considering a dark matter distribution with $BF=5$ (see section \ref{cps}) and $\rho$(local)=0.43 GeV/cm$^3$, annihilation cross sections of $\sigma v = 1.4 \times 10^{-25}\,\rm{cm^3/s}$, $3.6 \times 10^{-25}\,\rm{cm^3/s}$ and $7.6 \times 10^{-25}\,\rm{cm^3/s}$ are needed for these fits for 100, 300 and 600 GeV WIMPs, respectively. The solid line is the background-only prediction.}
\label{heatgauge}
\end{figure}

Next, considering annihilations primarily to gauge bosons, such as would be predicted for a higgsino or wino-like neutralino, we find that a reasonable fit to the HEAT data can be achieved for a WIMP with any mass above the gauge boson pair production threshold (although the fit improves for a heavier WIMP). In this case, we do not require the large diffusion constant used in figure~\ref{heatbbdif}. These results are shown in figure~\ref{heatgauge}.

\vspace{1.0cm}

\vspace{1.0cm}

\vspace{1.0cm}

\begin{figure}[thb]
\vbox{\kern2.4in\includegraphics{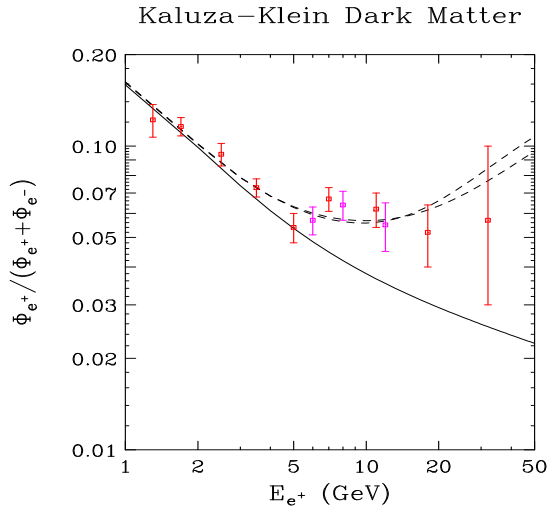}}
\caption{The positron fraction for Kaluza-Klein Dark Matter (KKDM) annihilations for WIMP masses of 300 and 600 GeV (top to bottom on the right side of the figure). In each case, the normalization of the dark matter contribution was chosen to maximize the quality of the fit to the HEAT data (shown as error bars). We find $\chi^2$'s per degree of freedom of 10.4/12 and 9.2/12 for 300 and 600 GeV WIMPs, respectively. Considering a dark matter distribution with $BF=5$ (see section \ref{cps}) and $\rho$(local)=0.43 GeV/cm$^3$, annihilation cross sections of $\sigma v = 5 \times 10^{-25}\,\rm{cm^3/s}$ and $1.2 \times 10^{-24}\,\rm{cm^3/s}$ are needed for these fits for 300 and 600 GeV WIMPs, respectively. The solid line is the background-only prediction.}
\label{heatKK}
\end{figure}

As with the case of annihilations to $b \bar{b}$, the cross section needed for annihilations to gauge bosons to fit the HEAT data are quite large, 1-8 $\times 10^{-25}$ cm$^3$/s for the mass range we studied with a boost factor of 5 and a local dark matter density of 0.43 GeV/cm$^3$. Although a thermal relic cannot have such a large annihilation cross section, it is interesting to note that this is roughly the range predicted for a neutralino LSP in the Anomaly Mediated Supersymmetry Breaking (AMSB) scenario with a non-thermal production mechanism.

Lastly, we show the fit of the positron spectrum from KKDM annihilations to the HEAT results in figure~\ref{heatKK}. For both masses we consider (300 and 600 GeV), we find an excellent fit to the data \cite{kkpos}. Again, the annihilation cross sections needed are quite large.

\section{Sensitivity of PAMELA and AMS-02}

Given the difficulty of a thermal dark matter relic to produce the positron flux observed by HEAT and the uncertainties regarding the effects of solar modulation at GeV energies, it is certainly possible that HEAT has not revealed a signature of dark matter. To pursue this question further, future experiments will be required to measure the positron spectrum with greater precision and at higher energies. The PAMELA and AMS-02 experiments will have the ability to perform such measurements.

In this section, we will not assume a positron spectrum consistent with the measurements of HEAT, but rather attempt to determine at what dark matter annihilation rate (or cross section) such a signature could be identified in future experiments.

At a minimum, for an experiment to claim the observation of a signature of dark matter annihilation, the spectral data must be statistically distinct from the predicted background spectrum. To evaluate this, we can calculate the $\chi^2$ of a given model for a set of data and compare it to the background only case.

To determine the $\chi^2$ of a set of data over the expected background, we perform the following sum:
\begin{equation}
\chi^2 = \sum \frac{(N_{\rm{Obs}}-N_{\rm{BG}})^2}{N_{\rm{Obs}}},
\label{chi2b}
\end{equation}
where the sum is over energy bins, $N_{\rm{Obs}}$ is the number of events observed in that bin and $N_{\rm{BG}}$ is the number of events predicted from the background contribution. Here we have assumed Gaussian errors. We have chosen to adopt energy bins of width $\Delta(\log E) = 0.60$ below 40 GeV and $\Delta(\log E) = 0.66$ above 40 GeV.

Just as we can convert the fluxes of figures~\ref{fluxbb}-\ref{fluxtautau} to positron fractions by using fluxes of background positrons and electrons, we can replace the fluxes (or numbers of events) used in Eq.~\ref{chi2b} with ratios of positrons to positrons+electrons to reduce the effects of solar modulation in our results. Because there are substantially more electrons than positrons observed, we can assume that there are negligible errors associated with the electron flux.

\begin{figure}[thb]
\vbox{\kern2.4in\includegraphics{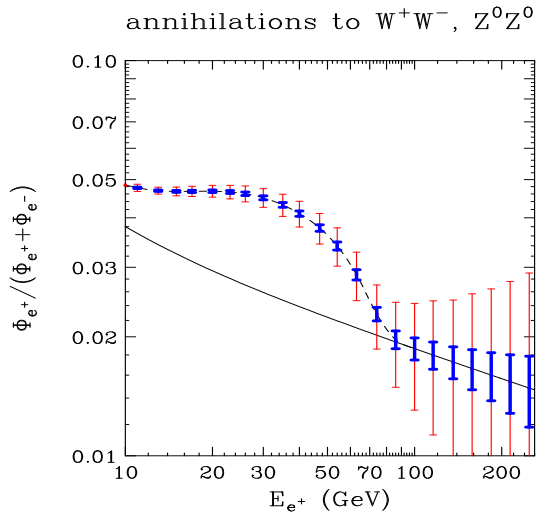}}
\caption{The positron fraction for dark matter annihilations to gauge bosons with an annihilation cross section of $10^{-25}\,\rm{cm^3/s}$, a boost factor (see section \ref{cps}) of 5, a local dark matter density of 0.43 GeV/cm$^3$ and a WIMP mass of 100 GeV. The light-red and dark-blue error bars shown are those projected for the PAMELA and AMS-02 experiments, respectively; each after three years of observations. The solid line is the background-only prediction.}
\label{stat1}
\end{figure}

For concreteness, we first consider the specific example of a 100 GeV WIMP annihilating to gauge bosons with a cross section of $10^{-25}\,\rm{cm^3/s}$, a boost factor of 5 and a local dark matter density of 0.43 GeV/cm$^3$. The positron fraction found for this scenario is shown in figure~\ref{stat1} along with the error bars after three years of observation by PAMELA (light-red) and AMS-02 (dark-blue). Following Eq.~\ref{chi2}, the total $\chi^2$ obtained for this flux is 1045.3 for PAMELA and 22,944 for AMS=02, both clearly distinguishable from the background. 

\begin{figure}[thb]
\vbox{\kern2.4in\includegraphics{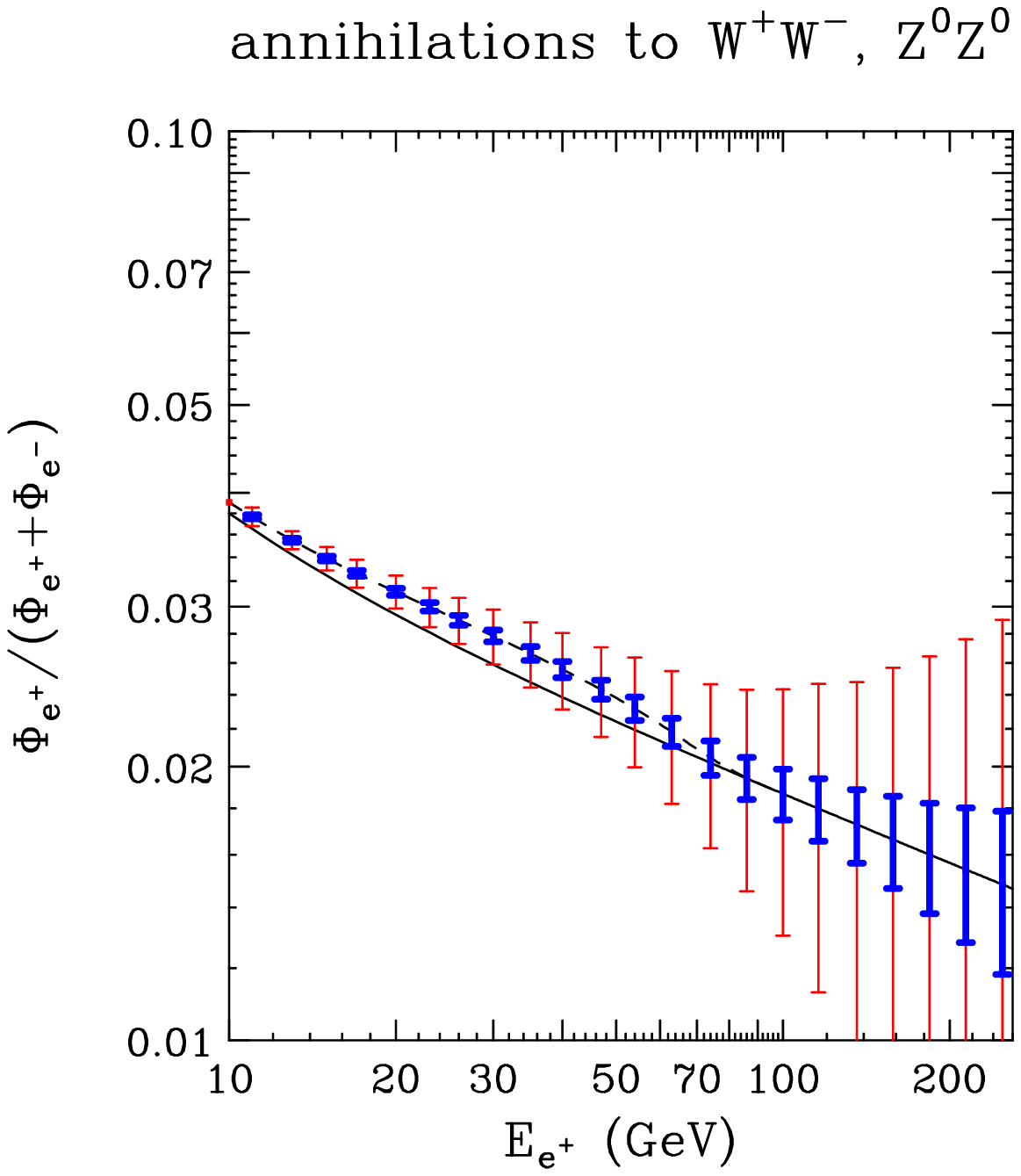}}
\caption{The positron fraction for dark matter annihilations with a WIMP with the same characteristics as in figure~\ref{stat1}, except with an annihilation cross section ten times smaller, $10^{-26}\,\rm{cm^3/s}$. Again, the light-red and dark-blue error bars shown are those projected for the PAMELA and AMS-02 experiments, respectively; each after three years of observations. The solid line is the background-only prediction.}
\label{stat2}
\end{figure}

If the annihilation rate is smaller, identifying such a feature will be more difficult. If we consider an annihilation rate ten times smaller than that used in figure~\ref{stat1}, we find the results shown in figure~\ref{stat2}. It is clear to the eye that the excess over background in this case is far less statistically significant. With three years of observation by PAMELA, the $\chi^2$ obtained is only 15.1. AMS-02, however, still obtains a $\chi^2$ of 1045.3.

\vspace{1.0cm}

\vspace{1.0cm}

\newpage

\vspace{1.0cm}

\begin{figure}[thb]
\vbox{\kern2.4in\includegraphics{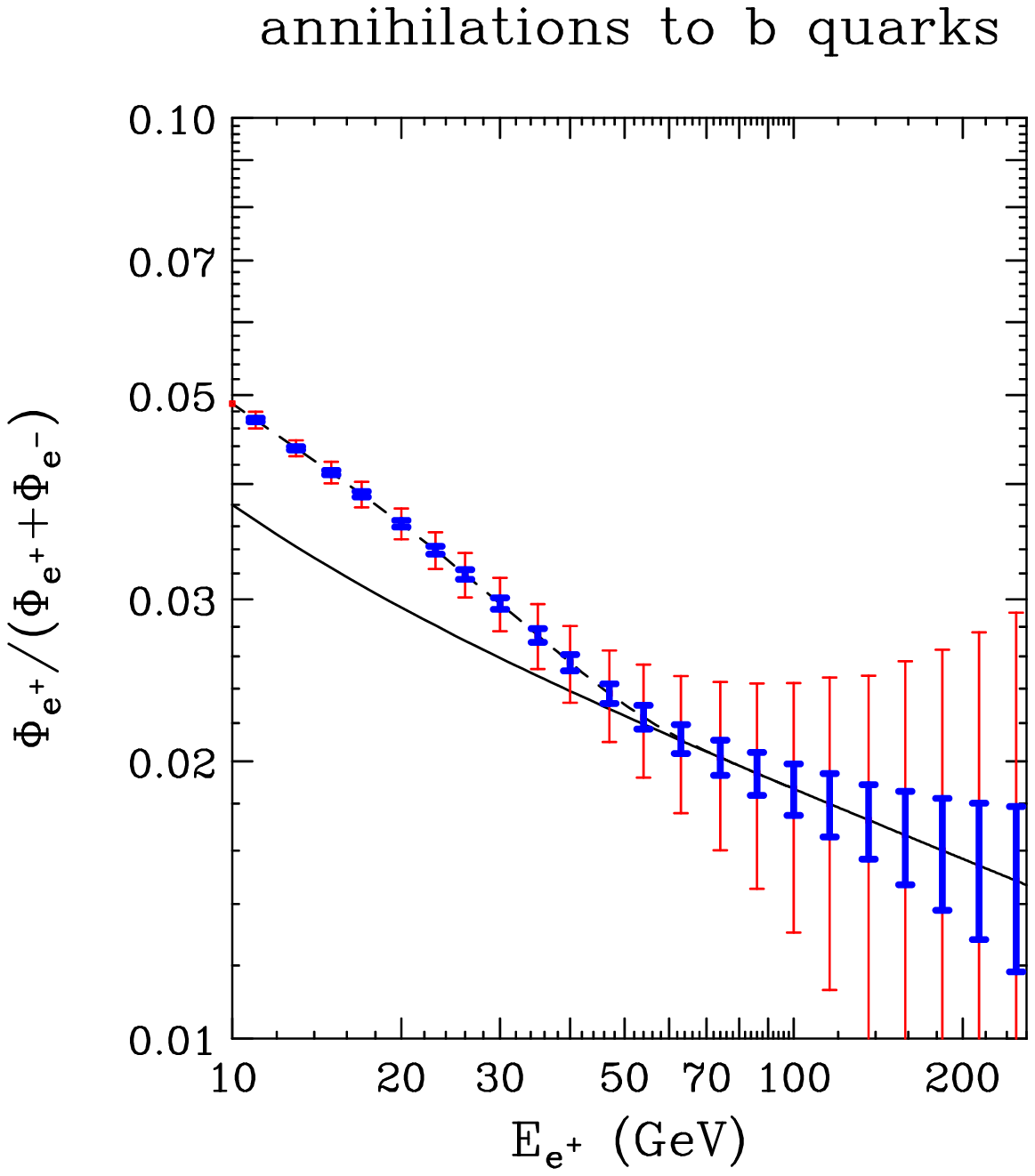}}
\caption{The positron fraction for dark matter annihilations to $b \bar{b}$ with an annihilation cross section of $10^{-25}\,\rm{cm^3/s}$, a boost factor (see section \ref{cps}) of 5, a local dark matter density of 0.43 GeV/cm$^3$ and a WIMP mass of 100 GeV. The light-red and dark-blue error bars shown are those projected for the PAMELA and AMS-02 experiments, respectively; each after three years of observations. The solid line is the background-only prediction.}
\label{stat3}
\end{figure}

\vspace{1.0cm}

\begin{figure}[thb]
\vbox{\kern2.4in\includegraphics{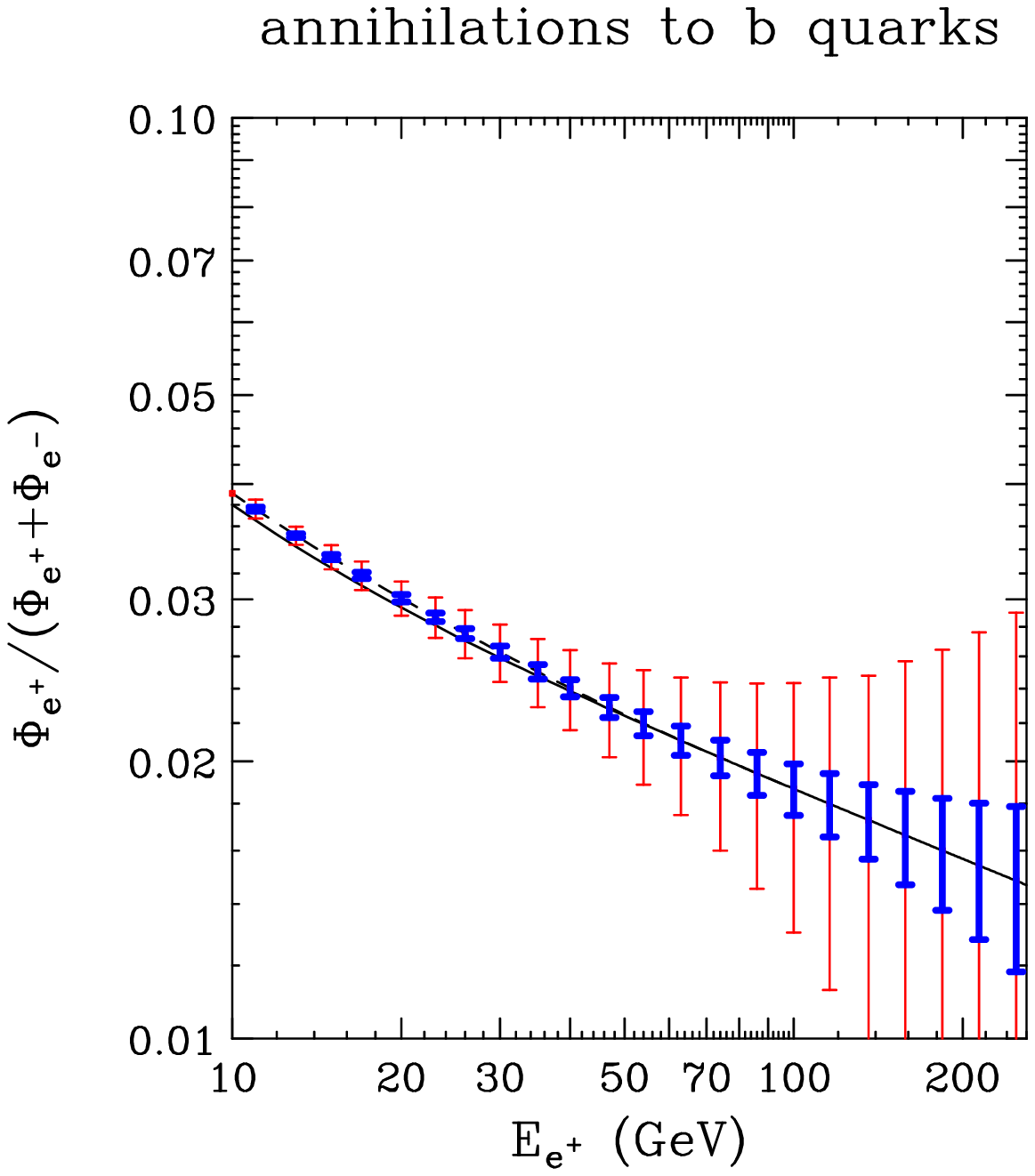}}
\caption{The positron fraction for dark matter annihilations with a WIMP with the same characteristics as in figure~\ref{stat3}, except with an annihilation cross section ten times smaller, $10^{-26}\,\rm{cm^3/s}$. Again, the light-red and dark-blue error bars shown are those projected for the PAMELA and AMS-02 experiments, respectively; each after three years of observations. The solid line is the background-only prediction.}
\label{stat4}
\end{figure}

\vspace{1.0cm}

\begin{figure}[thb]
\vbox{\kern2.4in\includegraphics{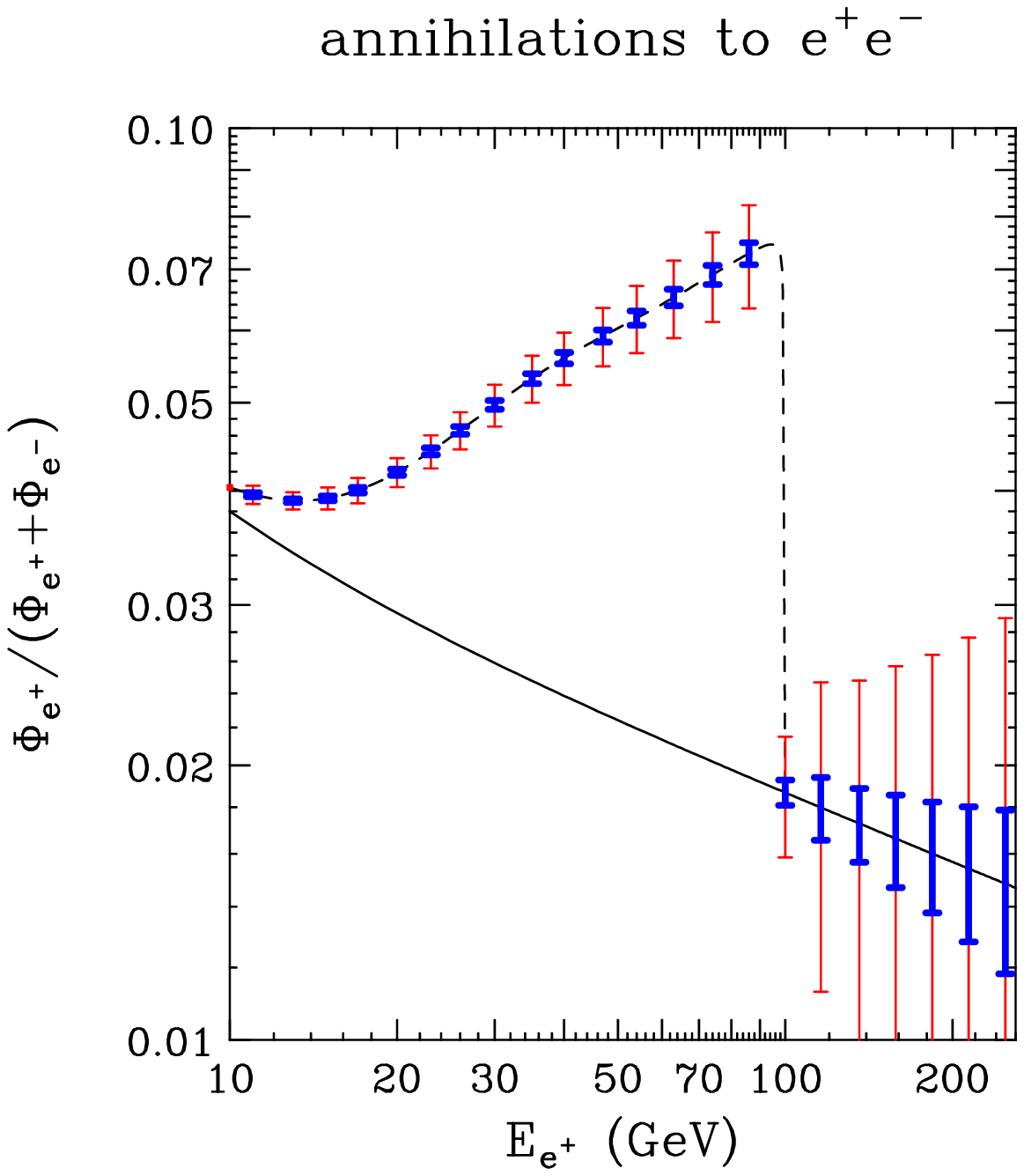}}
\caption{The positron fraction for dark matter annihilations to $e^+ e^-$ with an annihilation cross section of $10^{-26}\,\rm{cm^3/s}$, a boost factor (see section \ref{cps}) of 5, a local dark matter density of 0.43 GeV/cm$^3$ and a WIMP mass of 100 GeV. The light-red and dark-blue error bars shown are those projected for the PAMELA and AMS-02 experiments, respectively; each after three years of observations. The solid line is the background-only prediction.}
\label{stat5}
\end{figure}

\vspace{1.0cm}

\begin{figure}[thb]
\vbox{\kern2.4in\includegraphics{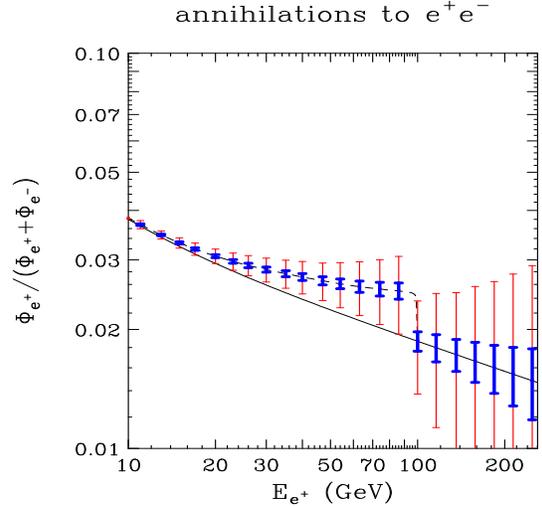}}
\caption{The positron fraction for dark matter annihilations with a WIMP with the same characteristics as in figure~\ref{stat5}, except with an annihilation cross section ten times smaller, $10^{-27}\,\rm{cm^3/s}$. Again, the light-red and dark-blue error bars shown are those projected for the PAMELA and AMS-02 experiments, respectively; each after three years of observations. The solid line is the background-only prediction.}
\label{stat6}
\end{figure}

\newpage

The situation is somewhat less optimistic for dark matter particles which annihilate to b quarks. These results are shown in figures~\ref{stat3} and~\ref{stat4}. Again, with the large cross section of $10^{-25}\,\rm{cm^3/s}$ (assuming $BF=5$, $\rho=0.43$ GeV/cm$^3$ and a 100 GeV WIMP), there is a significant excess over the expected background (see figure~\ref{stat3}). We calculate $\chi^2$'s of 9439 and 430 for AMS-02 and PAMELA, respectively. For the smaller cross section of $10^{-26}\,\rm{cm^3/s}$, the excess becomes very small and the $\chi^2$ falls to 5.4 and 118 at PAMELA and AMS-02 (see figure~\ref{stat4}), impossible to resolve with PAMELA and a challenge for AMS-02. 

The $\chi^2$ of the deviation from the background prediction is not the only measure of an experiment's ability to identify a contribution from dark matter annihilations. This becomes apparent when considering the case of dark matter which annihilates directly to positrons, {\it i.e.} $e^+ e^-$ pairs. The results for this case is shown in figures~\ref{stat5} for a cross section of $10^{-26}\,\rm{cm^3/s}$ and in figure~\ref{stat6} with $10^{-27}\,\rm{cm^3/s}$. Corresponding to figure~\ref{stat5}, we find $\chi^2$'s of 17,529 and 799 for AMS-02 and PAMELA, respectively. For the smaller cross section of figure~\ref{stat6}, we find $\chi^2$'s of 334 and 15.2. Note that these $\chi^2$'s are somewhat smaller than those corresponding to annihilations to gauge bosons (figures~\ref{stat1} and \ref{stat2}). Visually comparing these cases, it is obvious that the sudden discontinuity present in the $e^+ e^-$ case, but absent in the spectrum from annihilations to gauge bosons (or $b \bar{b}$, etc.), provides a clear signature of a new positron contribution which could not be incorporated into a modified background model. Despite the somewhat smaller $\chi^2$ values of the $e^+ e^-$ case, this spectrum has distinctive features which more than make up for this difference.

\begin{figure}[thb]
\vbox{\kern2.4in\includegraphics{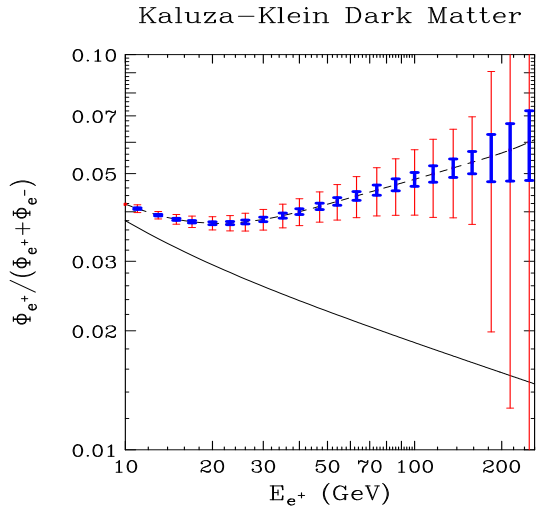}}
\caption{The positron fraction for Kaluza-Klein Dark Matter (KKDM) annihilations. A WIMP mass of 300 GeV, the minimum mass allowed for KKDM by electroweak precision measurements, was used. The annihilation rate shown corresponds to a boost factor (see section \ref{cps}) of 5, a local dark matter density of 0.43 GeV/cm$^3$ and an annihilation cross section of $10^{-25}\,\rm{cm^3/s}$.  The light-red and dark-blue error bars shown are those projected for the PAMELA and AMS-02 experiments, respectively; each after three years of observations. The solid line is the background-only prediction.}
\label{stat7}
\end{figure}

\begin{figure}[thb]
\vbox{\kern2.4in\includegraphics{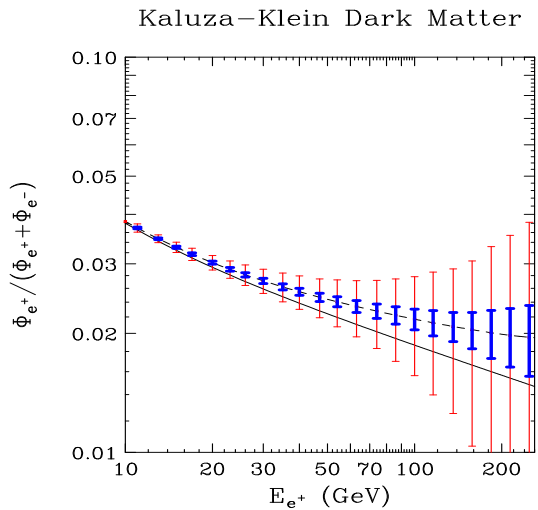}}
\caption{The positron fraction for dark matter annihilations with a WIMP with the same characteristics as in figure~\ref{stat7}, except with an annihilation rate ten times smaller. Again, the light-red and dark-blue error bars shown are those projected for the PAMELA and AMS-02 experiments, respectively; each after three years of observations. The solid line is the background-only prediction.}
\label{stat8}
\end{figure}

Lastly, our results for Kaluza-Klein Dark Matter (KKDM) are shown in figures~\ref{stat7} and \ref{stat8}. We calculate $\chi^2$ values of 393 and 8626 for the annihilation rate shown in figure~\ref{stat7}, for the PAMELA and AMS-02 experiments, respectively. For the smaller annihilation rate used in figure~\ref{stat8}, we find $\chi^2$'s of 5.9 and 129 for the two experiments. Note that in these two figures a mass of 300 GeV, corresponding to the minimum KKDM mass allowed by electroweak precision measurements, is used rather than the 100 GeV mass used in figures~\ref{stat1}-\ref{stat6}.

\vspace{1.0cm}

\vspace{1.0cm}

\newpage

\section{Reach of PAMELA and AMS-02}

To quantitatively assess the reach of PAMELA and AMS-02, we calculate the annihilation rate needed to be distinguished from the background at the 95\% confidence level. The 95\% confidence level corresponds to a $\chi^2$ per degree of freedom of approximately 1.3. For our method of energy binning, this corresponds to a total $\chi^2$ of about 30. Although this value can vary somewhat depending on the energy bins used and other assumptions, our over results are only slightly modified by these variations.

\begin{figure}[thb]
\vbox{\kern2.4in\includegraphics{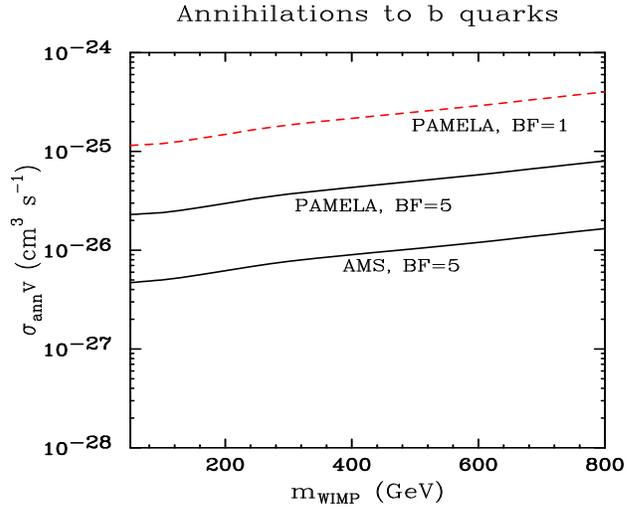}}
\caption{The ability of PAMELA and AMS-02 after three years of observation to detect positrons from dark matter annihilations into $b \bar{b}$ at the 95\% confidence level. Shown from top to bottom are the reach of PAMELA with a boost factor of 1 (dashed), PAMELA with a boost factor of 5 (solid) and AMS-02 with a boost factor of 5 (solid). Not shown is the reach of AMS-02 to dark matter with a boost factor of 1, which contour falls nearly on top of the PAMELA case for a boost factor of 5. Note that the maximum low velocity annihilation cross section for a thermal relic is approximately $3 \times 10^{-26}\,\rm{cm^3/s}$.}
\label{limitsbb}
\end{figure}

\begin{figure}[thb]
\vbox{\kern2.4in\includegraphics{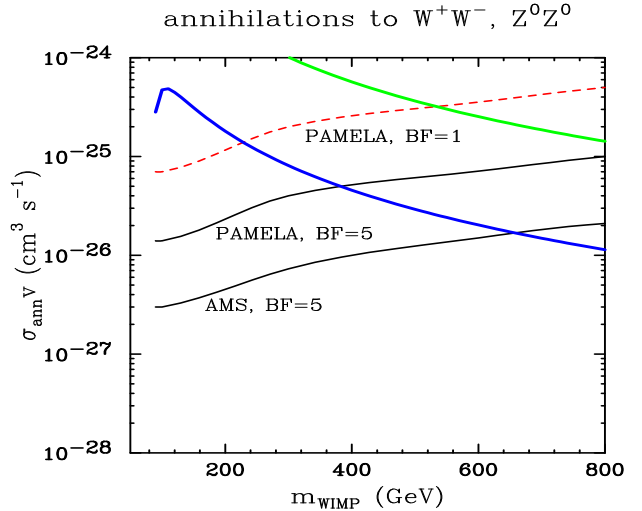}}
\caption{As in figure~\ref{limitsbb}, except for dark matter which annihilates into gauge bosons. Also shown as thick solid lines are the predicted annihilation cross sections for a pure higgsino-LSP (lower, blue) and a nearly pure wino-LSP in the AMSB scenario (upper, green). See section~\ref{neudm} for more details.}
\label{limitsgauge}
\end{figure}

\begin{figure}[thb]
\vbox{\kern2.4in\includegraphics{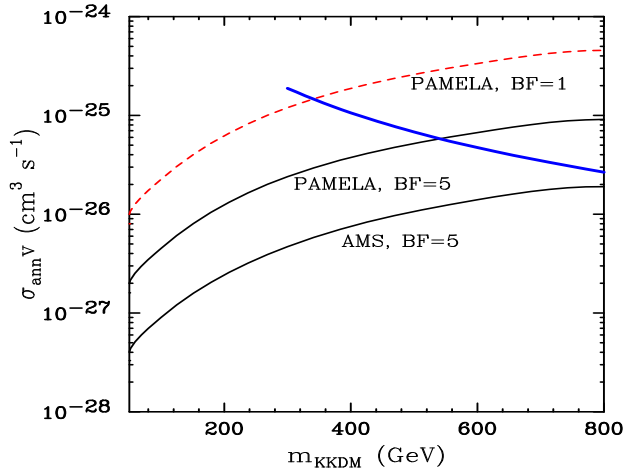}}
\caption{As in figure~\ref{limitsbb}, except for Kaluza-Klein Dark Matter (KKDM), annihilating in the modes described in section~\ref{kkdmsec}. Shown as a solid blue line is the annihilation cross section predicted for KKDM. PAMELA will be able to test for the presence of KKDM up to masses of 350-550 GeV, while AMS-02 will test up to masses of 550-1000 GeV.}
\label{limitsKK}
\end{figure}

For a given WIMP mass and annihilation mode(s), we can find the dark matter annihilation rate that corresponds to the 95\% confidence level sensitivity. The annihilation rate can be further separated into the low velocity annihilation cross section and the Boost Factor (BF) (see section~\ref{cps}). Throughout this section, we assume a mean local dark matter of $\rho=0.43$ GeV/cm$^3$. The annihilation rates which are observable in three years of observation by PAMELA or AMS-02 are shown in figures~\ref{limitsbb}, \ref{limitsgauge} and \ref{limitsKK}.

For a bino-like, neutralino dark matter candidate, we see the range of annihilation cross sections which are assessable to PAMELA and AMS-02 in figure~\ref{limitsbb}. With typical boost factors of a few, PAMELA may be capable of observing a bino-like neutralino with a cross section near the maximum allowed for a thermal relic (a few times $10^{-26}\,\rm{cm^3/s}$). The prospects for detection are particularly good for models with large values of $\tan \beta$ (see figures~\ref{sigma5} and \ref{sigma50}). For AMS-02, the sensitivity is considerably improved, reaching cross sections around $10^{-27}\,\rm{cm^3/s}$. We find very similar results if we consider a WIMP which annihilates 90\% to $b \bar{b}$ and 10\% $\tau^+ \tau^-$ or 50\% to $b \bar{b}$ and 50\% $t \bar{t}$ (see section~\ref{neudm}). 

For a higgsino or wino-like LSP, our results are shown in figure~\ref{limitsgauge}. Also shown in this figure are the predicted annihilation cross sections to gauge boson pairs for a pure higgsino and a pure wino within the Anomaly Mediated Supersymmetry Breaking (AMSB) scenario. Due to the nearly degenerate chargino present in the AMSB scenario, LSP annihilation cross sections to gauge bosons are very large, as can be seen in the figure. After three years, PAMELA will be able of testing the AMSB scenario with masses as large as 550 GeV and 1 TeV, for boost factors in the range of 1 to 5. AMS-02 will be sensitive to this scenario for masses up to 2-3 TeV. For a higgsino, PAMELA will be sensitive up to 230-380 GeV and while AMS-02 can probe as high as 400-650 GeV. If a mixed gaugino-higgsino is the LSP, results somewhere between those shown in figures~\ref{limitsbb} and \ref{limitsgauge} will apply.

\vspace{1.0cm}

\vspace{1.0cm}

\newpage

\vspace{1.0cm}

Our results for Kaluza-Klein Dark Matter (KKDM) are shown in figure~\ref{limitsKK}. We find that PAMELA will test this scenario for masses up to 350-550. AMS-02 will reach up to 1 TeV.

\section{Comparisons to Other Detection Methods}

In this section, we will briefly compare the prospects for observing particle dark matter with cosmic positrons to dark matter detection using other techniques. 

The prospects for direct detection experiments \cite{direct}, which look for the effects of WIMPs elastically scattering off of an detector, are difficult to compare to those of cosmic positron experiments. This is because direct detection experiments largely depend on the elastic scattering cross section of a WIMP with the target material, whereas the success of positron experiments depends on the WIMP's annihilation cross section (at low velocities). These two cross sections are not directly determined by each other, and are very difficult to compare in any general way. 

The prospects for dark matter detection with neutrino telescopes (by observing neutrinos produced in the annihilations of WIMPs trapped in the Sun or Earth) \cite{indirectneutrino} are also difficult to compare to positron experiments. Again, these experiments largely depend on the elastic scattering cross section of a WIMP with nucleons rather than a WIMP's annihilation rate.

Searches for gamma-rays produced in WIMP annihilations \cite{indirectgamma} is the method of indirect detection which has been studied in the greatest detail. The prospects for this method depend strongly on the distribution of dark matter in our Galaxy, particularly in the Galactic Center. If a high density of dark matter (a cusp or a spike) is present in the Galactic Center, it has been shown that future gamma-ray experiments such as GLAST or HESS would likely be able to detect the annihilation signal. Very recently, however, HESS has announced the detection of a bright gamma-ray source at the Galactic Center which may interfere with the future detection of gamma-rays produced in dark matter annihilations \cite{hesssource}. This HESS source has a spectrum consistent with a power law extending at least to several TeV. Although it has been shown that this, in principle, could be the product of dark matter annihilation, it would require an exceedingly heavy WIMP (above 12 TeV), which is probably less likely than the presence of a astrophysical source at this location. With such a source coincident with the dark matter annihilation region at the Galactic Center, the continuum component of the dark matter gamma-ray flux will be difficult to identify, leaving only the much more faint line emission to be observed in future gamma-ray experiments.

In addition to positrons, other anti-matter species can be used to search for evidence of dark matter annihilations. In particular, observations of cosmic anti-protons \cite{antiprotons,ullio} and anti-deuterons \cite{ullio,Edsjo:2004pf} may be capable of revealing signatures of particle dark matter. Measurements of the cosmic anti-proton spectrum have excellent prospects for dark matter detection, although they depend critically on unknown parameters of the Galactic halo, so are difficult to assess with much confidence. In particular, the width and extent of the galactic diffusion zone can dramatically alter the rates and spectra predicted for anti-protons. Positrons, on the other hand, lose energy much more rapidly and therefore sample only the surrounding few kpc of the dark matter halo, thus making the diffusion zone boundary conditions less important. Anti-deuteron studies certainly look promising, although less attention has been given to these than to other detection channels thus far.

\section{Conclusions}

Future measurements of the cosmic positron spectrum represent one of the most promising methods for detecting particle dark matter. The PAMELA and AMS-02 experiments each represent major steps forward in the precision measurement of the cosmic positron spectrum, for the first time measuring up to hundreds of GeV. If dark matter consists of weakly interacting particles present in our Galactic halo, it is fairly likely that these experiments will be sensitive to it. 

If the dark matter is a supersymmetric neutralino, the prospects for its detection by PAMELA and AMS-02 depend on its composition. If it is a nearly pure higgsino (superpartner of the neutral Higgs bosons), PAMELA will be sensitive to masses up to 230 to 380 GeV, after three years of observation, depending on the degree of local inhomogeneities in the dark matter distribution. AMS-02 will be sensitive up to 400 to 650 GeV for a higgsino LSP. If the LSP is bino-like, as is the case in much of the parameter space of the Minimal Supersymmetric Standard Model (MSSM), the prospects for these experiments depend strongly on the WIMP's annihilation cross section (at low velocities). PAMELA will be sensitive to cross sections up to the order of a few times $10^{-26}$ cm$^3$/s while AMS-02 can improve on this by a factor of about four. Supersymmetric models with large values of $\tan \beta$ are particularly likely to fall near or above this range. Thirdly, if supersymmetry comes in the form of Anomaly Mediated Supersymetry Breaking (AMSB), the wino LSP will be quite observable by both PAMELA and AMS-02, probing such models up to multi-TeV masses.

If the dark matter consists of Kaluza-Klein excitations of Standard Model fields (the excitation of the hypercharge gauge boson, in particular), their annihilations should produce an observable positron flux to AMS-02, and possibly to PAMELA.  PAMELA will be sensitive to KKDM masses up to 350 to 550 GeV, depending on the degree of local inhomogeneities in the dark matter distribution. AMS-02 will be sensitive up to 600 GeV to 1 TeV.

The prospects for the detection of WIMPs from their positron signatures with PAMELA and AMS-02 are quite encouraging. We look forward to the deployment of these experiments in the coming years.

\vspace{0.5cm}

{\it Acknowledgments}: We would like to thank Graham Kribs, Mirko Boezio and Piero Ullio for helpful comments and discussions. DH is supported by the Leverhulme Trust.

\end{document}